\documentclass{iopart}
\usepackage{epsf}
\usepackage{iopams}
\usepackage{graphicx, wrapfig}
\usepackage{array}
\usepackage{amssymb}
\usepackage{bm}

\newcommand{\beq}{\begin{equation}} 
\newcommand{\eeq}{\end{equation}} 
\newcommand{\beqn}{\begin{eqnarray}} 
\newcommand{\eeqn}{\end{eqnarray}} 
\newcommand{\beaa}{\begin{eqnarray*}}
\newcommand{\eeaa}{\end{eqnarray*}}
\newcommand{\dis}{$\displaystyle}
\newcommand{\pa}{\partial}

\newcommand{\gabd}{g_{\alpha\beta}}

\newcommand{\gmabd}{\gamma_{ab}}
\newcommand{\gmaa}{\gamma_a\!{}^\alpha}
\newcommand{\gmbb}{\gamma_b{}^\beta}

\newcommand{\tgmabu}{\tilde\gamma^{ab}}
\newcommand{\tgmabd}{\tilde\gamma_{ab}}
\newcommand{\tgamma}{\tilde\gamma}
\newcommand{\gmcdu}{\gamma^{cd}}

\newcommand{\Tabd}{T_{\alpha\beta}}

\newcommand{\Gabd}{G_{\alpha\beta}}

\newcommand{\Sabd}{S_{ab}}
\newcommand{\albe}{{\alpha\beta}}

\newcommand{\tR}{{}^{3}\!R}
\newcommand{\ttR}{{}^{3}\!\tilde R}
\newcommand{\Rnl}{R^{\rm NL}}

\newcommand{\tA}{\tilde A}

\newcommand{\Kabd}{K_{ab}}

\newcommand{\tAabd}{\tilde A_{ab}}
\newcommand{\zeroD}{{\raise1.0ex\hbox{${}^{\ \circ}$}}\!\!\!\!\!D}
\newcommand{\zD}{{\raise1.0ex\hbox{${}^{\ \circ}$}}\!\!\!\!\!D}
\newcommand{\zLap}{{\raise1.0ex\hbox{${}^{\ \circ}$}}\!\!\!\!\Delta}

\newcommand{\Lie}{\mbox{\pounds}}

\newcommand{\nalam}{\mathrel{\raise0.9ex\hbox{$^\lambda$}\mkern-14mu
\lower0.0ex\hbox{$\nabla$}}}

\newcommand{\compa}{(M/R)_\infty}
\def\agt{\mathrel{\raise.3ex\hbox{$>$}\mkern-14mu\lower0.6ex\hbox{$\sim$}}}
\def\alt{\mathrel{\raise.3ex\hbox{$<$}\mkern-14mu\lower0.6ex\hbox{$\sim$}}}

\begin{document}

\title{Models of helically symmetric binary systems} 
 
\author{Shin'ichirou Yoshida$^{1,4}$, Benjamin C. Bromley$^2$, Jocelyn S. Read$^1$, K\=oji Ury\=u$^{1,3}$,
and John L. Friedman$^1$}

\address{
$^1$ 
Department of Physics, University of Wisconsin-Milwaukee, P.O. Box 413,  
Milwaukee, WI 53201, U.S.
}
\address{
$^2$
Department of Physics, University of Utah, Salt Lake City, Utah 84112, U.S.
}
\address{
$^3$ 
SISSA, via Beirut 4, 34014 Trieste, Italy
} 
\address{
$^4$ 
Department of Physics, Florida Atlantic University, Boca Raton FL
33431, U.S.
} 
\ead{jsread@uwm.edu}
%
%
%
%


\begin{abstract}
Results from helically symmetric scalar field models and first results from 
a convergent helically symmetric binary neutron star code are reported here; 
these are models stationary in the rotating frame of a source with constant 
angular velocity $\Omega$.  In the scalar field models and the neutron star 
code, helical symmetry leads to a system of mixed elliptic-hyperbolic character.
The scalar field models involve nonlinear terms of the form $\psi^3$, $(\nabla\psi)^2$, 
and $\psi\Box\psi$ that mimic nonlinear terms of the Einstein equation. Convergence 
is strikingly different for different signs of each nonlinear term; it is typically 
insensitive to the iterative method used; and it improves with an outer 
boundary in the near zone.  In the neutron star code, one has no control on the 
sign of the source, and convergence has been achieved only for an outer boundary less 
than $\sim 1$ wavelength from the source or for a code that imposes helical symmetry only inside 
a near zone of that size.  The inaccuracy of helically symmetric solutions with appropriate 
boundary conditions should be comparable to the inaccuracy of a waveless formalism
that neglects gravitational waves; and the (near zone) solutions we obtain for 
waveless and helically symmetric BNS codes with the same boundary conditions nearly 
coincide.

\end{abstract} 

\maketitle  
 
\section{Introduction} 
\label{intro} 

Initial data for the inspiral of binary neutron star (BNS) systems 
and corresponding quasiequilibrium BNS models have been based 
either on the initial value equations alone or on the IWM (Isenberg-Wilson-Mathews) 
spatially conformally flat ansatz.  In each case one solves a truncated 
version of the Einstein equation for a metric having fewer than 
the six independent potentials that remain after a choice of gauge. 
The error of the approximation limits the accuracy of the first 
orbits of simulations in two ways: The initial data has spurious radiation; 
and, more importantly, the balance between gravitational attraction and orbital 
acceleration is not enforced, leading to orbits that are not exactly circular.

One way to go beyond spatial conformal flatness is to construct 
an analog in the full theory of the Newtonian binaries that are stationary 
in a rotating frame.  In general relativity, these models are helically 
symmetric spacetimes \cite{fus02,ggb02}, 
with equal amounts of ingoing and outgoing radiation.  Binary black holes of 
this kind were first discussed by Blackburn and Detweiler \cite{bd92},
and models involving nonlinear scalar wave equations have been studied by 
a group of researchers organized by Price (henceforth the {\em Consortium})
\cite{whelan00,whelan02,price04,andrade04,torre03,bromley05}.  Because 
the power radiated by a helically symmetric 
binary is constant in time, the spacetime cannot be asymptotically flat.  
At distances large compared to $1/\Omega$, however, the spacetime of a 
binary system with a helical Killing vector approximates asymptotic flatness 
--  until, beyond about $10^4 M$ for neutron-star models of mass $M$, 
the enclosed energy in gravitational waves is comparable to the mass of the binary.  
The approximation is similar to, and possibly more accurate than, a 3rd 
post-Newtonian approximation in which the 2 1/2 post-Newtonian
radiation is omitted. 

In this paper, we report the construction of a convergent 
neutron-star code in which the full Einstein equation, together with the 
equation of hydrostatic equilibrium, is solved numerically under the assumption 
of helical symmetry.  Convergence relies on a boundary that is not much 
farther than a wavelength from the system, and we present results from a 
number of related nonlinear scalar field models in which convergence 
requires either a small coefficient of the nonlinear term or a boundary 
close to the source.  The results of these toy models are surprising in two 
ways.  First, convergence of the scalar-field models is most 
strongly affected by the sign of the source term, with one choice of sign 
yielding a convergent solution for remarkably large values of the nonlinear 
terms we examined.  Second, convergence does not depend strongly on the iterative 
method used to solve the equation, on whether one uses, for example, a Newton-Raphson 
iteration or an iteration based on a Green function that inverts only a 
convenient part of the second-order nonlinear operator.  

The plan of the paper is as follows.
Sect. 2 introduces the set of toy scalar field models with nonlinear 
terms chosen to mimic the nonlinear terms in the dynamical part of the 
Einstein equation.  We describe several iteration methods, one closely related 
to that of the Uryu code, the others to codes developed by Andrade \textit{et. al.} in
\cite{andrade04} and further by Bromley, Owen, and Price \cite{bromley05}.  
Sect. \ref{sec:NumAcc} compares the accuracy of codes based on the various iteration methods.
Sect. \ref{sec:RConv} reports the main results on convergence of the scalar-field models.
Finally, Sect. \ref{sec:bns} describes Uryu's neutron-star code, presents its first results, 
and compares the solution it yields to that of a closely related code based on a 
waveless formalism \cite{WAT05}. 

\section{Toy Problems}

\subsection{Helically symmetric, nonlinear scalar wave equations}
We consider a scalar field $\psi$ on Minkowski space, satisfying a wave equation 
with a source $s$ that mimics two objects in circular orbit and with a 
nonlinear term $\cal N[\psi]$, whose strength is adjusted by a coefficient $\lambda$:  
\begin{equation}
\Box\psi - \lambda \mathcal{N}[\psi] = s.
\label{toy1}\end{equation}
We use three different nonlinear terms: $\mathcal{N}[\psi] = \psi^3$, 
$\mathcal{N}[\psi]=|\nabla\psi|^2$ with $\nabla$ the spatial gradient, and 
$\mathcal{N}[\psi]=\psi\Box\psi$, chosen to represent the types of nonlinear 
terms that appear in dynamical components of the field equations.  

The source $s$ is a sum of two 3-dimensional gaussian distributions, 
\beq
s(t,r,\theta,\phi) = \sum_\pm\frac{q}{\sqrt{(2 \pi)^{3}}} \
\exp
\left(-\frac{\left( \vec{r} \pm \vec{R}(t)\right)^2}{\sigma^2} \right),
\eeq
centered about points $\pm \bf R$, 
$\displaystyle {\bf{R}}(t) = a \left[ \cos(\Omega t)\hat{\bf x} + \sin(\Omega t)\hat{\bf y} \right]$, 
each a distance $a$ from the origin and each having spread $\sigma^2$ and total charge $q$. 
The source $s$ is stationary in a frame moving with 
angular velocity $\Omega$; that is, it is Lie-derived by the helical 
Killing vector 
\beq k^\alpha = t^\alpha + \Omega \phi^\alpha \eeq
of Minkowski space, where $t^\alpha$ and $\phi^\alpha$ (equivalently $\partial_t$ and $\partial_\phi$)
are generators of time-translations and of rotations in the plane of the binary source. 

One can regard the scalar-field models as toy models of neutron stars of mass $M$, if the charge $q$ of each gaussian source is identified with $4\pi M$.  In gravitational units
(G=c=1), all quantities of a binary star system can be specified in terms of $M$.  In the models presented below, we set $q=1$, $a=1$, $\sigma=0.5$, and $\Omega=0.3$, corresponding
to a binary system of mass $M$, stellar separation $2a=8\pi M$, stellar radius 
$\sigma = 2\pi M$, and velocity $v=a\Omega = 0.3$.

A helically symmetric solution, like a genuinely stationary solution, is given 
by its value on a spacelike slice and the field equation (\ref{toy1}) can be 
written in a form that involves only spatial derivatives.  That is, using 
the symmetry relation \dis \Lie_k\psi = (\partial_t +\Omega\partial_\phi)\psi =0$  
to replace time derivatives by $\phi$ derivatives, one 
can rewrite Eq.~(\ref{toy1}) at $t=0$ in the spatial form,  
\begin{equation} \label{waveq}
(\nabla^2 - \Omega^2 \partial^2_\phi)\Psi
- \lambda\mathcal{N}[\Psi] = S, 
\end{equation} 
where \dis \Psi(r,\theta,\phi )=\psi(t=0,r,\theta,\phi), \  
S(r,\theta,\phi) = s(t=0,r,\theta,\phi)$. 
Then \dis \psi(t,r,\theta,\phi) = \Psi(r,\theta,\phi - \Omega t)$.

The operator \dis {\cal L} := \nabla^2 - \Omega^2 \partial^2_\phi$ has a mixed character, 
elliptic inside the light cylinder $\Omega\sqrt{x^2+y^2} = 1$, hyperbolic outside.  
The difficulties in finding numerical solutions stem from this behavior.  
In finding an iterative 
solution, one inverts the operator $\cal L$, but $\cal L$ is not negative, and it 
lacks the contraction property that underlies the convergence of iterative schemes 
used to invert nonlinear elliptic equations and to prove existence of exact solutions. 

\subsection{Numerical methods} \label{sec:Methods} 

\noindent{\em KEH method}

In the KEH method one splits off the linear, flat-space operator $\cal L$ and 
inverts it by a choice ${\cal L}^{-1}$ of Green function.  
The iterative solution, beginning with $\Psi_0 = {\cal L}^{-1} S$, is then given by
\begin{equation} \label{lineq}
\Psi_{n+1} = \Psi_0 + \lambda {\cal L}^{-1}{\cal N}[\Psi_n]  
\label{iter}\end{equation}

Although $\cal L$ is not elliptic,  the operator associated with each spherical 
harmonic is the elliptic Helmholtz operator: 
\dis {\cal L}[\Psi_{lm}(r)Y_{lm}] = [\nabla^2 +m^2\Omega^2][\Psi_{lm}(r)Y_{lm}]$. The 
corresponding radial operator \dis \frac{d^2}{dr^2} +  \frac{2}{r}\frac{d}{dr} 
- \frac{l (l+1)}{r^2}  + m^2 \Omega^2$ has as its eigenfunctions the spherical Bessel 
functions, from which a Green function is constructed.    
For definiteness, we choose as ${\cal L}^{-1}$ the form that, for a bounded source, 
yields the half-advanced+half-retarded solution, namely 
\begin{equation}
{\cal L}^{-1}S := \sum_{lm}\int dr' S_{lm}(r') g_{lm}(r,r')Y_{lm}(\theta,\phi), 
\label{eq:green}\end{equation}
with 
\begin{equation}
g_{l0} = \frac{1}{2l+1}\frac{r_<^l}{r_>^{l+1}}, 
\qquad
g_{lm} = m \Omega j_l(m\Omega r_<) n_l(m\Omega r_>), \ m\ne 0.
\end{equation}
Here $r_< = \mathrm{min}(r,r')$, $r_> = \mathrm{max}(r,r')$, 
and $j_l(x)$ and $n_l(x)$ are the spherical Bessel functions of the first and second kinds.

At each iteration, the nonlinear term ${\cal N}[\Psi_n]$ serves as an effective source.
The polynomial nonlinear function $\Psi^3$ is most easily computed by
shifting from $\Psi_{lm}(r)$ back into $\Psi(r,\theta,\phi)$ and cubing at
each point, while the derivative-based nonlinear terms, $|\nabla\Psi|^2$ and
$\Psi\Box\Psi$, are calculated using the properties of the spherical
harmonics.

Finally, as is usual in codes to solve nonlinear elliptic equations, we use softening 
and continuation to extend the range of convergence to larger values of $\lambda$.
That is, instead of using $\Psi_{n+1}$ as defined in Eq.~(\ref{iter}), we can use a
softened $\Psi_{n+1}^\omega$ defined by
\begin{equation}
\Psi_{n+1}^\omega = \omega \Psi_{n+1} + (1 - \omega)\Psi_{n}.
\end{equation}
Given a converged field solution for some nonlinearity with small
$\lambda$, it is sometimes possible to use continuation to obtain a
solution for larger $\lambda$: The converged solution to Eq.~(\ref{waveq})
with weak nonlinearity is used as the initial field $\Psi_0$ for the
iteration of Eq.~(\ref{iter}) for strong nonlinearity. In this way one 
moves along a sequence of solutions with increasing values of $\lambda$.
The effectiveness of softening and convergence is explored in
Section \ref{sec:RConv}.

\noindent{\em Finite difference and eigenspectral methods}

The Finite Difference code uses an iteration based on the Newton-Raphson method,
with numerical approximations that reduce it in part to a secant method. 
Write the equation to be solved as
\begin{equation}
F =  {\cal L}\Psi - \lambda\mathcal{N}[\Psi] - S = 0.
\end{equation}
Numerically, $\Psi$ is given by a set of values $\Psi_i$ on the
three-dimensional spatial slice. Given an initial field $\Psi_i$, each
iterative step generates a modification $\delta\Psi_i$ by inverting 
\begin{equation}
J_{ij}\delta\Psi_j = - F_i
\end{equation}
where $J_{ij}$ is the Jacobian
\begin{equation}
J_{ij} = \frac{\partial F_i}{\partial\Psi_j}
\end{equation}

The Helmholtz operator has the form $({\cal L}\Psi)_i = L_{ij} \Psi_j$ where
$L_{ij}$ is constructed from finite difference operations and incorporates 
boundary conditions. The corresponding part of
$J_{ij}$ is simple.

\begin{eqnarray}
J_{ij} & = & \frac{\partial}{\partial \Psi_j}\left[ 
		L_{ik}\Psi_k - \lambda ( \mathcal{N}[\Psi])_i - S_i \right] \\
	& = & L_{ij} - \lambda \frac{\partial \mathcal{N}[\Psi]_i}{\partial \Psi_j}
\end{eqnarray}
The nonlinear piece of the Jacobian is evaluated numerically by
varying local field values.

The Eigenspectral code \cite{bromley05} uses the same iterative scheme as the Finite
Difference method, but it employs adapted
coordinates and a discretized spectral decomposition. 
To specify the adapted coordinate system, we begin with comoving
Cartesian coordinates ($\tilde{x},\tilde{y},\tilde{z}$) where the
$\tilde{z}$-axis is the axis of rotation. The axes are rotated to a set
\begin{equation}
\tilde{X}=\tilde{y},\tilde{Y}=\tilde{z},\tilde{Z}=\tilde{x}
\end{equation}
for which the $\tilde{Z}$-axis is a line through the center of each source.
The adapted coordinates are chosen to approach spherical polar
coordinates, with $\Theta$ measured from the $\tilde{Z}$ axis, far 
from the sources.
For each point ($\tilde{X},\tilde{Y},\tilde{Z}$) , let $r_1$ and $r_2$ be the distances from the source
centers, $\theta_1$ and $\theta_2$ corresponding angles from the $\tilde{Z}$
axis.  Then, 
\begin{eqnarray}
r_1 & = & \sqrt{(\tilde{Z} - a)^2 + \tilde{X}^2 + \tilde{Y}^2},\\
r_2 & = & \sqrt{(\tilde{Z} + a)^2 + \tilde{X}^2 + \tilde{Y}^2}.
\end{eqnarray}
The adapted coordinates are
\begin{eqnarray}
\chi & = & \sqrt{r_1 r_2}\\
\Theta & = & \frac{1}{2}(\theta_1 + \theta_2)
	= \frac{1}{2} \tan^{-1}\left(
		\frac{2 \tilde{Z} \sqrt{\tilde{X}^2 + \tilde{Y}^2}}
			{\tilde{Z}^2 - a^2 - \tilde{X}^2 - \tilde{Y}^2}\right) \\
\Phi &=& \tan^{-1}(\tilde{X}/\tilde{Y})  
\end{eqnarray}

The spectral decomposition involves the angular Laplacian of $\Theta$ and 
$\Phi$, 
\begin{equation}
D^2\Phi :=\frac{1}{\sin\Theta}
\frac{\partial}{\partial\Theta}
\left[\sin\Theta \frac{\partial}{\partial\Theta}\right]
+
\frac{1}{(\sin\Theta)^2}\frac{\partial^2}{\partial\Phi^2};
\end{equation}
note that $D^2$ is {\em not} the angular part of $\nabla^2$ in adapted coordinates, 
but it agrees asymptotically with the usual angular Laplacian far from the source.
Instead of the spherical harmonics of the continuum Laplacian $D^2$, the eigenspectral 
code uses the exact eigenvectors of the matrix $L$ obtained by discretizing $D^2$ 
on the adapted coordinate grid $\{\Theta_i,\Phi_j\}$. Angular derivatives are represented in $L$ by 
second order finite differencing.  That is, with 
$
\left[\sin\Theta D^2 \Phi\right](\Theta_a,\Phi_b) 
\mbox{ approximated by } \sum_{ij} L_{ab,ij} \Psi(\Theta_i,\Phi_j)$,

the normalized eigenvectors $Y_{ij}^{(k)}$ of $L$ satisfy
\beq
\sum_{ij} L_{ab,ij} Y_{ij}^{(k)} = 
- \Lambda^{(k)} \sin \Theta_a Y_{ab}^{(k)}, 
\eeq
\beq
\sum_{ij} Y_{ij}^{(k)}Y_{ij}^{(k')}\sin \Theta_i \Delta\Theta
\Delta\Phi = \delta_{k k'}.
\eeq

With the field expanded in terms of the eigenvectors $Y_{ij}^{(k)}$,
\begin{equation}
\Psi(\chi,\Theta_i,\Phi_j) = \sum_k a^{(k)}(\chi) Y_{ij}^{(k)},
\end{equation}
and the operator $\cal L$ written in terms of adapted coordinates 
(Eq. 8-17 of \cite{bromley05}), ${\cal L}\Psi$ is expressed in terms 
of the $Y_{ij}^{(k)}$ (Eq. 31 of \cite{bromley05}), 
\begin{equation}
\sum_{k} \left( \alpha_{k',k} \frac{d^2 a^{(k)}(\chi)}{d \chi^2}
+ \beta_{k',k} a^{(k)}(\chi) + \gamma_{k',k}\frac{d a^{(k)}(\chi)}{d \chi}\right) =
S_{k'}\ ,
\end{equation}
where the $\alpha_{k',k}, \beta_{k',k}$, and $\gamma_{k',k}$ involve angular
derivatives of the $Y^{(k)}$ computed by finite differencing
(Eqs. 41-43 of \cite{bromley05}) . This equation for $a^{(k)}$ is iterated in the 
same fashion as in the Finite Difference method to find the solution with nonlinear terms.

If all harmonics were retained, the Eigenspectral method would
be the equivalent of the Finite Difference method in adapted coordinates,
although in a different basis. The advantage of the adapted coordinates is 
that the distribution of points encodes most of the physically relevant
information in the low-order harmonics. Its disadvantage is that higher 
order harmonics require an increasingly cumbersome formalism.  The code is 
consequently limited to harmonics $l\leq 2$, saving enough memory to allow 
high resolution in the radial coordinate.

\subsection{Boundary conditions} 

The KEH method at each iteration finds a solution that has standing wave
behaviour for the flat space piece, imposing boundary conditions by the
choice of Green function. There remains freedom to add a homogeneous solution at
each iteration, and this has been used to study the sensitivity of convergence 
to the choice of boundary condition. 

The Eigenspectral and Finite Difference methods include boundary conditions
in the finite difference matrix for the linear $\Box$ operator. Outgoing
conditions are imposed on the edges of the grid by enforcing 
\begin{equation}
(\partial_r \Psi - \partial_\phi \Psi)_{r=r_{\rm max}} = 0
\end{equation}
A solution for ingoing radiation can then be generated by a spatial
inversion across the plane through the sources and perpendicular to their
rotation. At each step, a periodic solution is constructed by superposition
of the ingoing and outgoing solutions.

\section{Estimating numerical accuracy} \label{sec:NumAcc}
\subsection{KEH code} \label{sec:NumKEH}

Several types of numerical approximation in the KEH code produce effects
that can be estimated by convergence testing. Most obvious is the choice
of spatial grid in $(r,\theta,\phi)$ on which numerical integration is
performed. Very high resolution in $\theta$ and $\phi$ is easily obtained.
The number of radial points is more problematic with our simple equispaced
grids. A Richardson extrapolation error estimate from varying radial grid
spacing shows that precision of $10^{-5}$ is estimated for runs comparing
code results. Estimation of the range of $\lambda$ giving convergent
solutions is done at lower radial precision.

\begin{figure}
\begin{center}
\begin{tabular}{cc}
Linear ($\mathcal{N}[\Psi] =0 $) & $\mathcal{N}[\Psi] = \Psi^3$ \\
\includegraphics[height=40mm]{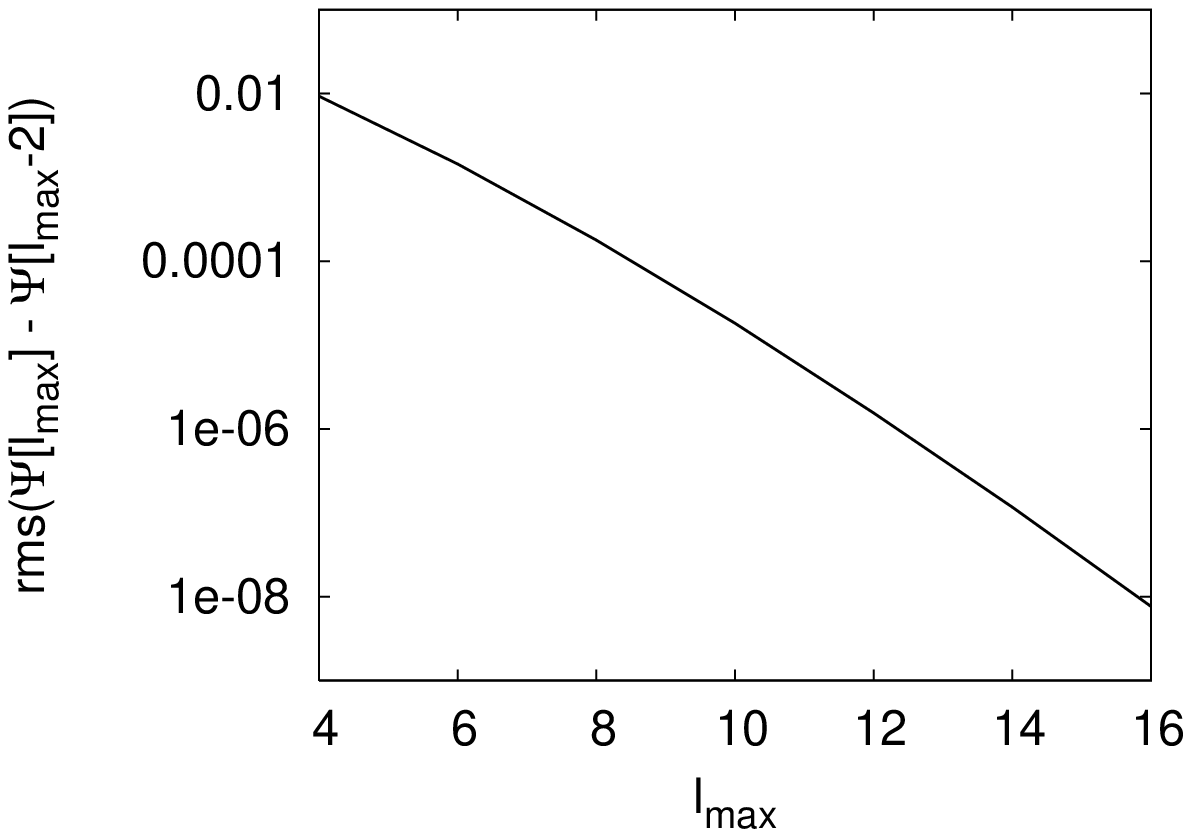} &
\includegraphics[height=40mm]{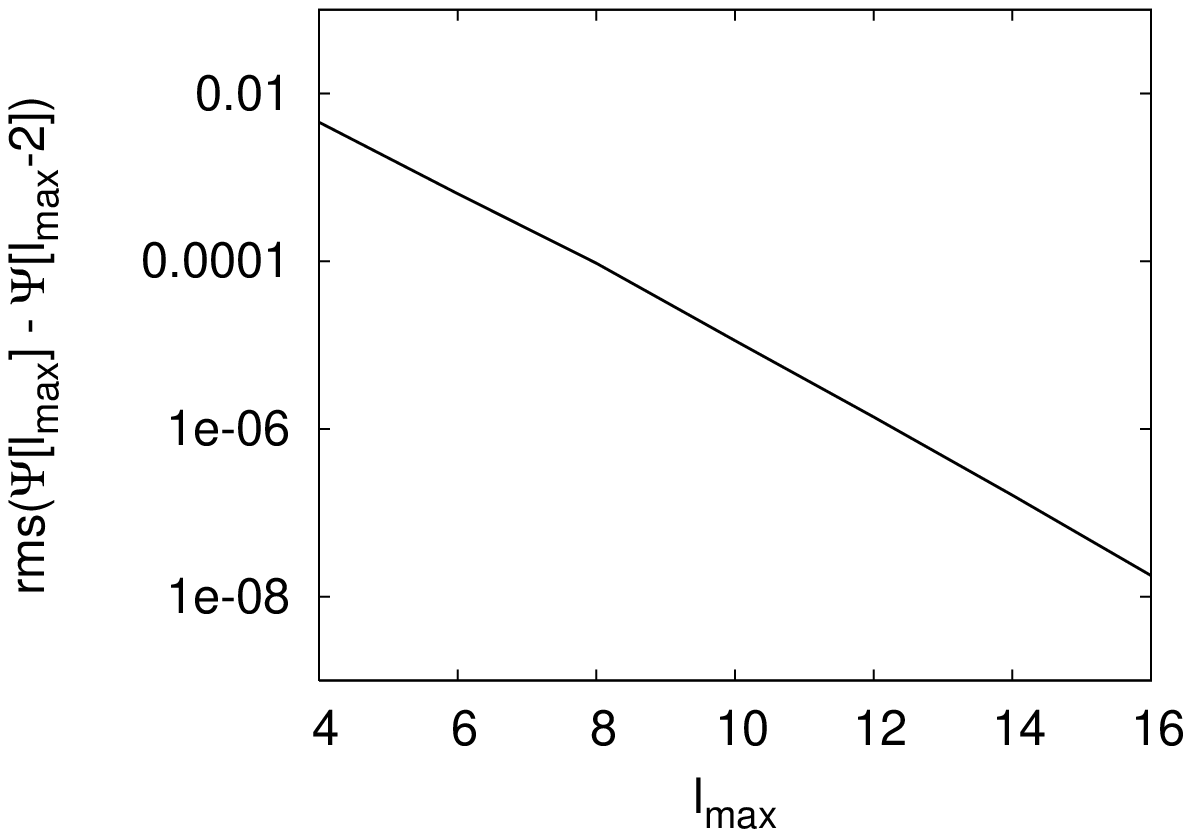} \\
$\mathcal{N}[\Psi] = |\nabla\Psi|^2$ & $\mathcal{N}[\Psi] = \Psi\Box\Psi$ \\
\includegraphics[height=40mm]{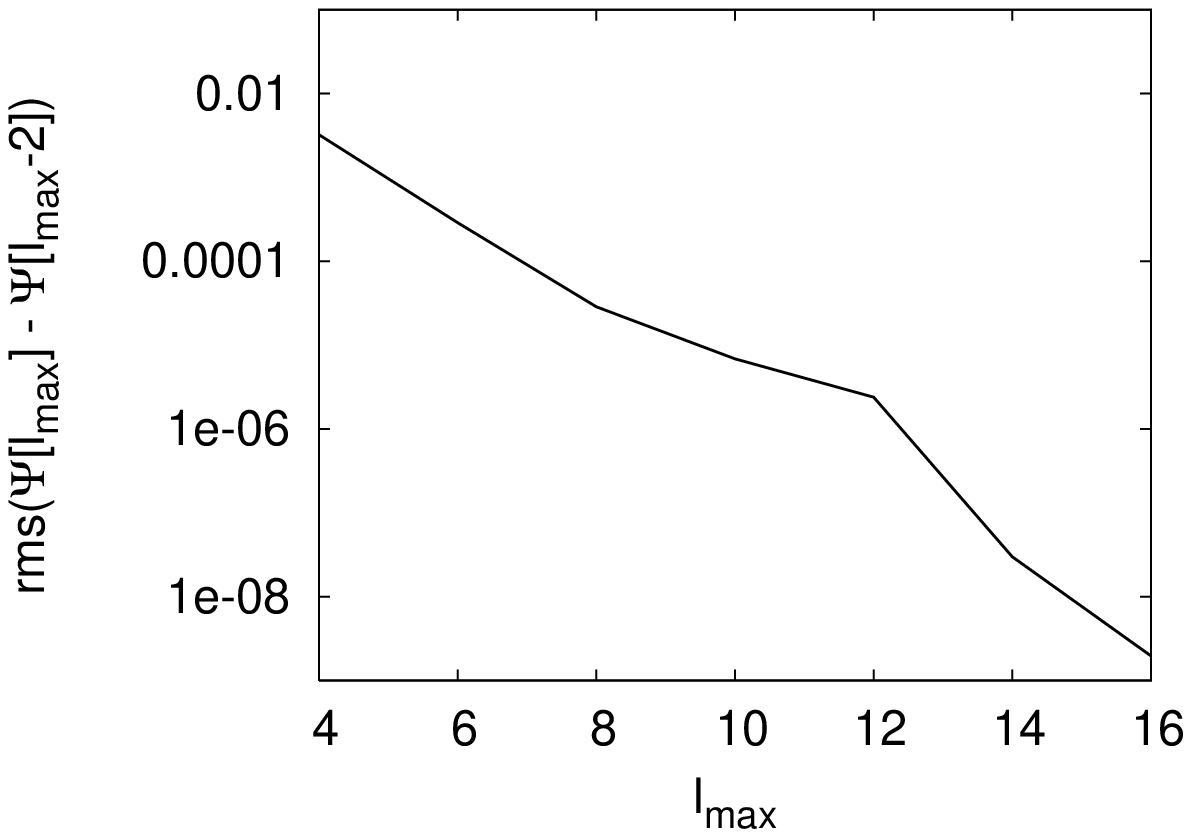} &
\includegraphics[height=40mm]{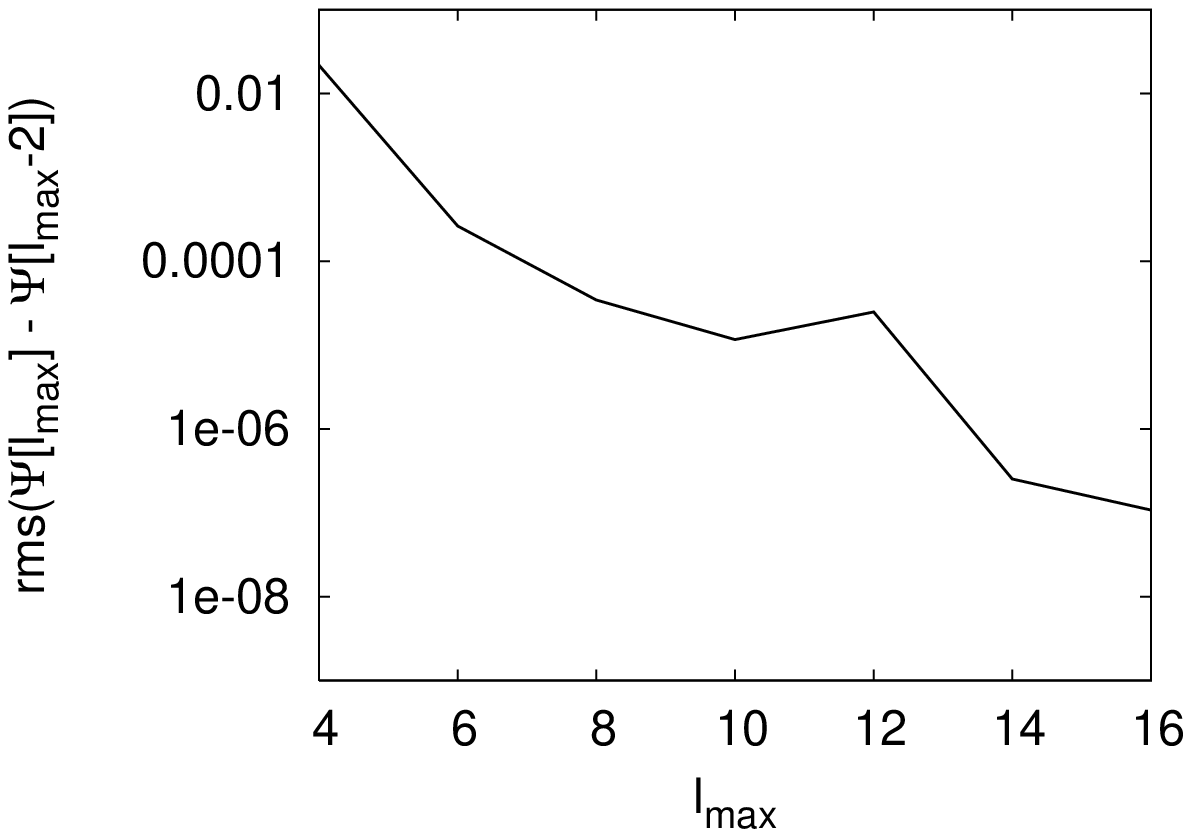}
\end{tabular}	
\caption{The rms change in the converged field as the
number of harmonics considered increases.}
\label{fig:linharm}
\end{center}
\end{figure}

As the KEH method rests on the decomposition of the field into spherical
harmonics, accuracy will depend on the number of harmonics retained in the
numerical calculation.  Examining the difference between results with an
increasing number of harmonics retained shows that a good approximation is to
use up to $l=12$ in the code.  The fractional difference between the field calculated with
$l_{\rm max}=12$ and the field including higher harmonics
is less than $10^{-6}$ at each point.   Figure \ref{fig:linharm} displays the rms difference in solutions as the number of harmonics retained by the code is increased.

In the toy models presented below, a solution is computed at each iteration using the half-advanced+half-retarded 
Green function (\ref{eq:green}).  Because the linear field $\Psi$ of a perpetually radiating source 
falls off like $r^{-1}$, the nonlinear terms ${\cal N}= (\nabla\Psi)^2$ and  
${\cal N}= \Psi\Box\Psi$ that serve as effective sources for each iteration do not fall off fast enough 
for the integrals to converge, if the outer boundary extends to infinity. One can, however, pick out a solution
that is independent of the outer boundary by fixing the value of $\Psi$ at a finite radius $R$.  That is, one 
can, at each iteration, add the homogeneous solution that maintains the specified value of $\Psi$ at $R$. 
This has remarkably little effect on the field in the region close to
the sources, with less that a $1\%$ change in field strength for points 
with $r<6$.  This insensitivity of the near-zone field to the amplitude of the waves, when 
the source dominates the solution, is the reason a helically symmetric solution makes sense 
as an approximation to an outgoing solution.    

\subsection{Comparing codes}\label{sec:NumComp}

With different 3D grid patterns for the codes, it is most
straightforward to compare results on rays through the volume of
interest. We compare the results extrapolated to three orthogonal axes: 
taking the $x$-axis through the centers of the two sources and the 
$z$-axis perpendicular to the plane of rotation.  Both $x$ and $y$ axes 
show wave behaviour away from the sources; along the $z$ axis, $\psi$ shows only 
Coulomb-type behavior, because $Y_{lm}$ vanishes on the axis when $m\neq 0$, 
and for $m=0$, ${\cal L}\psi = \nabla^2 \psi$.    

\begin{figure}[htp]
\begin{center}
\begin{tabular}{c c}
Linear ($\mathcal{N}[\Psi] =0 $) & 
$\mathcal{N}[\Psi] = \Psi\Box\Psi, \lambda= -0.7$ \\
\includegraphics[height=42mm]{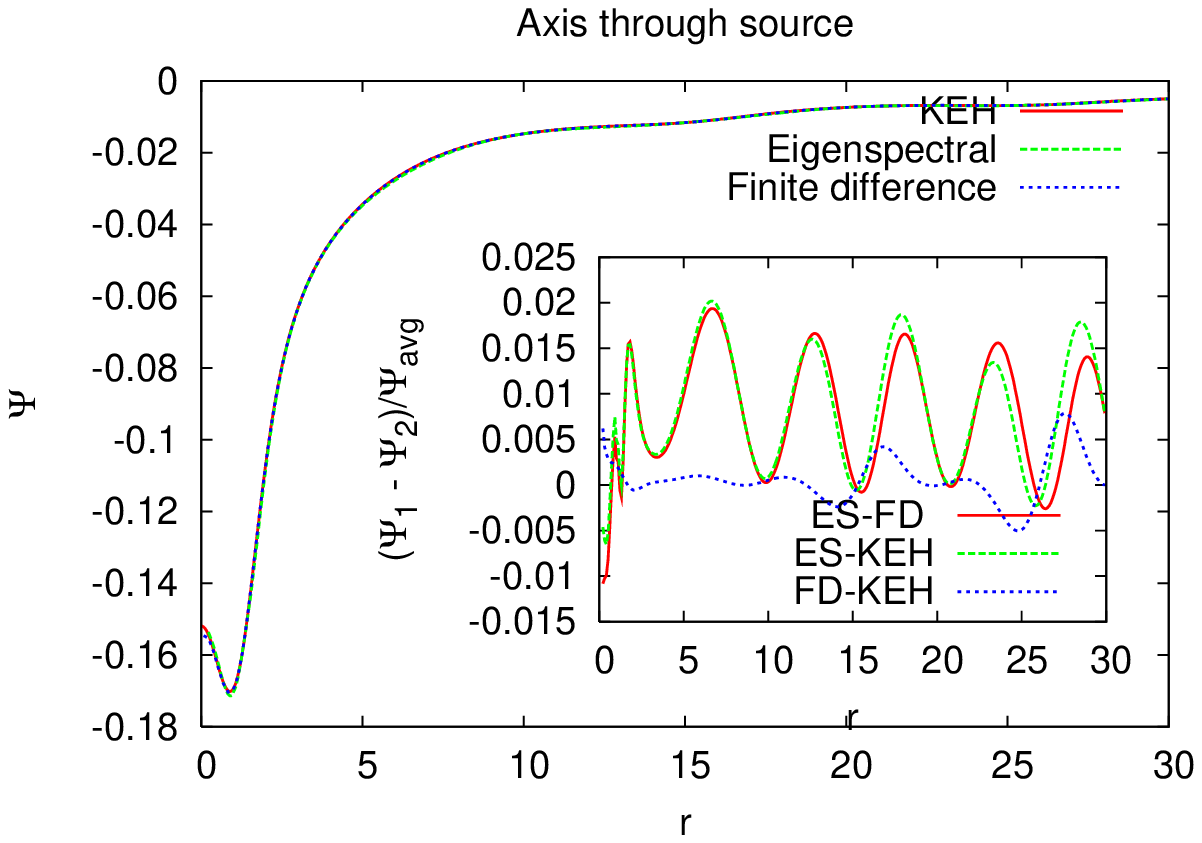} &
\includegraphics[height=42mm]{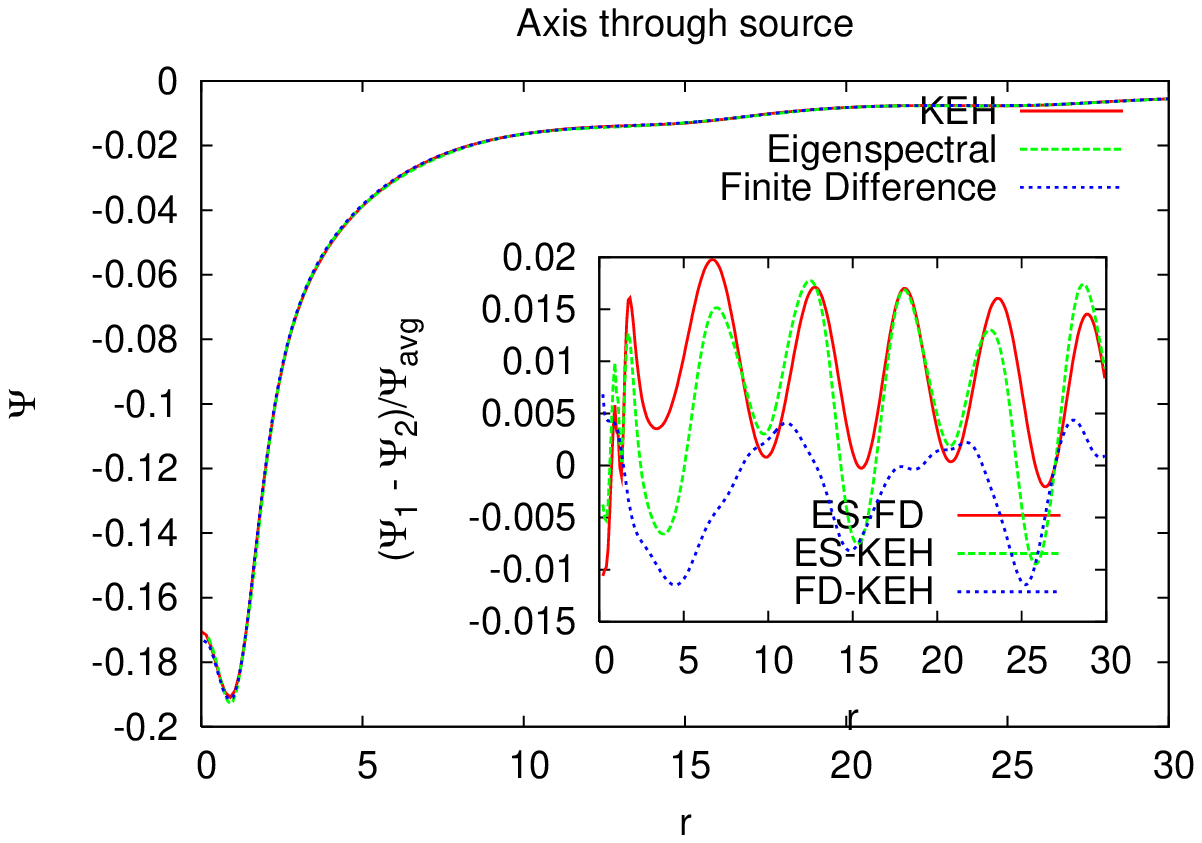}\\
$\mathcal{N}[\Psi] = \Psi^3, \lambda = 100 $ & 
$\mathcal{N}[\Psi] = \Psi^3, \lambda= -2.2$ \\
\includegraphics[height=42mm]{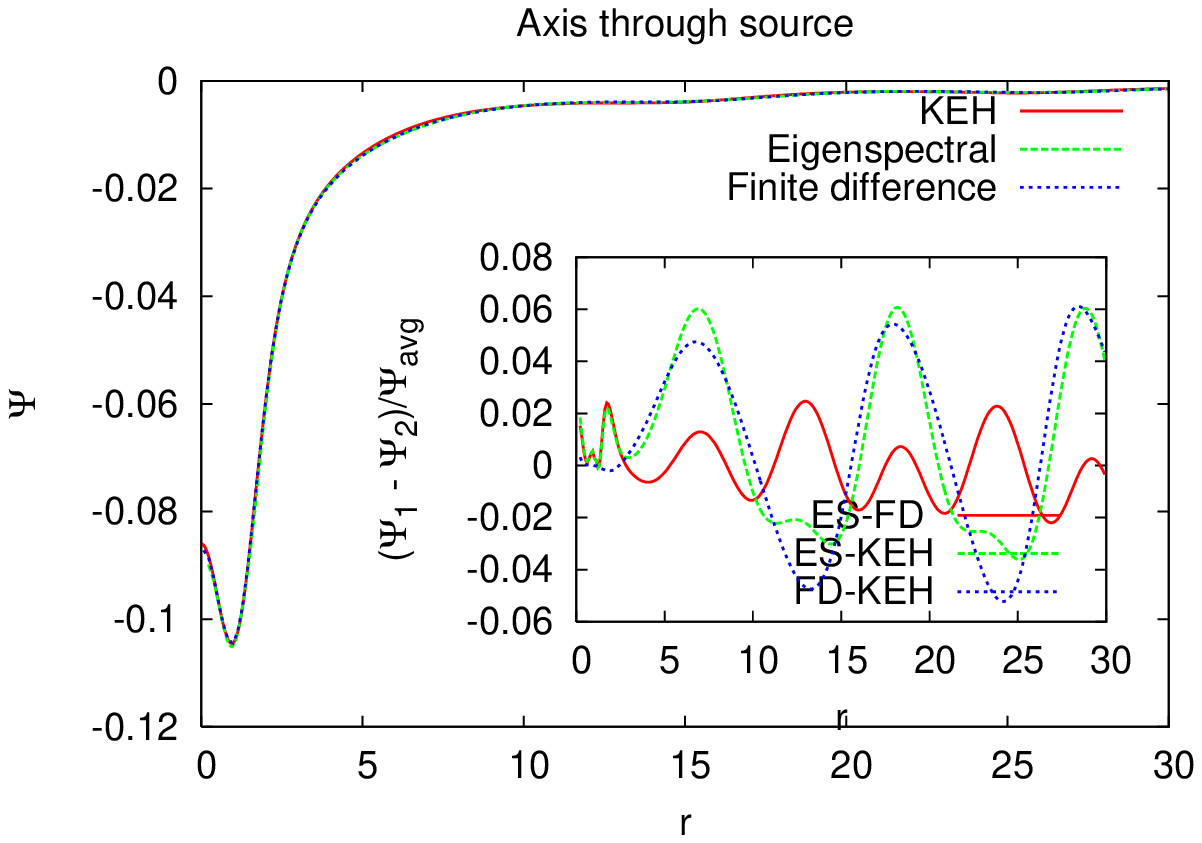} &
\includegraphics[height=42mm]{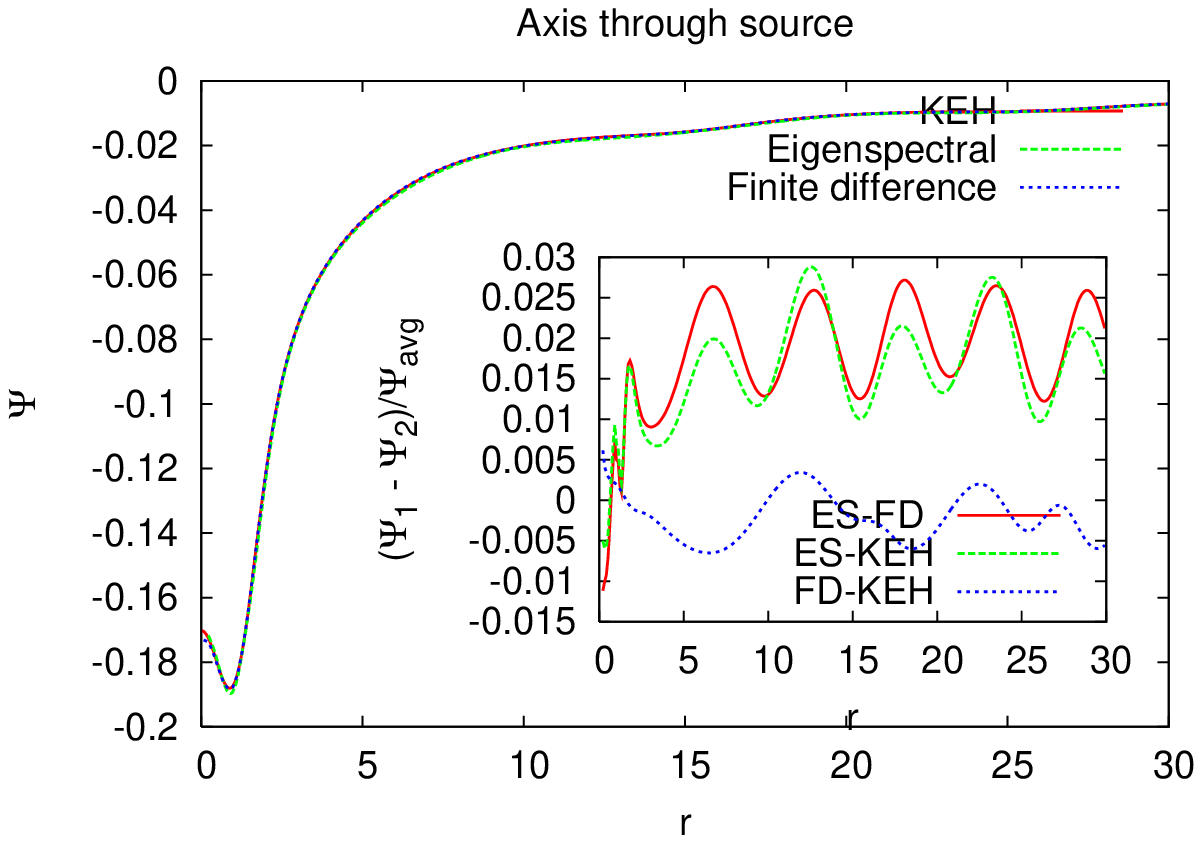} \\
$\mathcal{N}[\Psi] = |\nabla\Psi|^2, \lambda=3$ &
$\mathcal{N}[\Psi] = |\nabla\Psi|^2, \lambda=-100$ \\
\includegraphics[height=42mm]{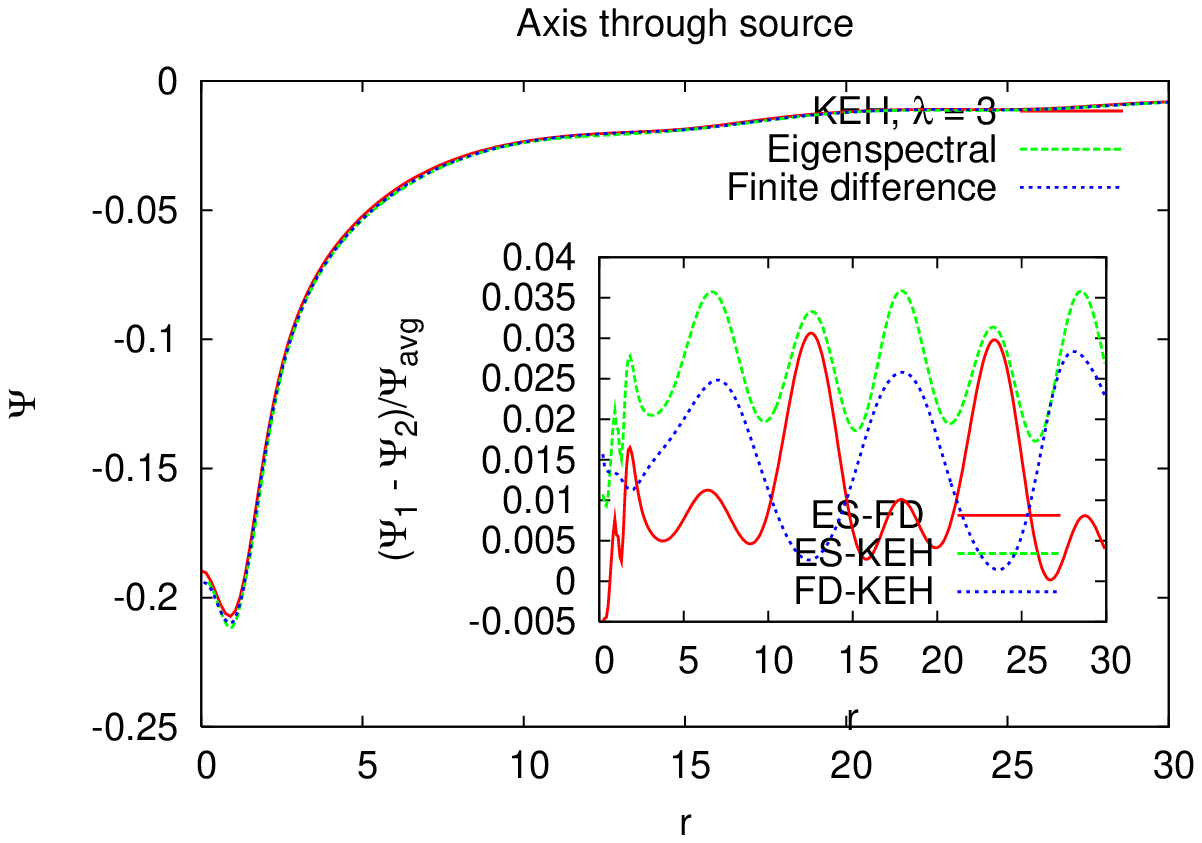} &
\includegraphics[height=42mm]{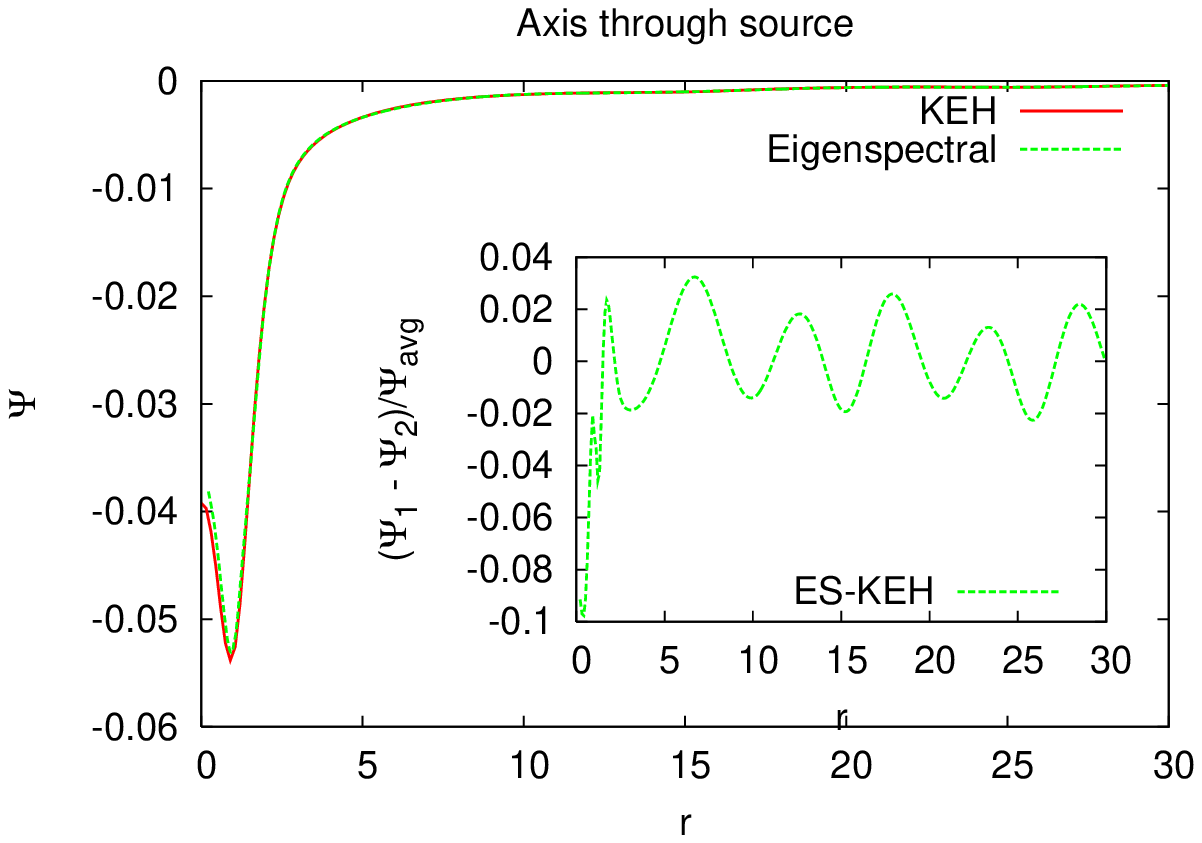} \\
\end{tabular}
\caption{The scalar field $\Psi$ as a function of distance $r$ from
the origin in units of the orbital radii along the source axis.  
Each panel corresponds to a model specified by the form of the 
nonlinear term $\mathcal{N}[\Psi]$ and the value of the parameter $\lambda$, as
written above the panels.  In all cases, the angular frequency of rotation is
$\Omega = 0.3$ and the source strengths are unity.  The plots show results from
the KEH method (solid red curves), the Eigenspectral method (dashed green) 
and the Finite Difference code (dotted blue). The insets give more detailed
comparisons between these results; the curves show the difference
between pairs of solutions relative to the average field.
The Eigenspectral method gives a stronger periodic modulation
relative to the other methods, due to its restriction to low-order
harmonics. 
} 
\label{fig:comparecode}
\end{center}
\end{figure}

Sample comparison plots are shown in Figure \ref{fig:comparecode}. Further comparisons 
are shown in the longer version of this paper on the gr-qc archive. 

The field values on the axes are interpolated for comparison.  
Differences between fields are divided by the average field value of the three 
codes to find relative error, as plotted in the insets of Figure 
\ref{fig:comparecode}. We compute the rms of this relative error for the grid points 
along each axis.   These rms values are expressed as percentages in Table \ref{table:rmsdiff}. In the worst case, the rms difference is $~3\%$ between codes.

A smaller rms difference on the rotation axis than on the
source and perpendicular axes indicates that discrepancies in the wave region
dominate, as in the FD-KEH comparison with $\mathcal{N}[\Psi] = \Psi^3$,
$\lambda = 100 $. In other cases, the error on all three axes is comparable,
and there is some shift between codes seen even on the rotation axis, as in the
FD-KEH comparison with $\mathcal{N}[\Psi] = |\nabla\Psi|^2$, $\lambda = 3 $. 

\begin{table}[htp]
\caption{A comparison of code output for scalar models. The values in
the table give the percent rms difference between of $\Psi$-values
from three numerical methods as in Figure~1.  The rms differences were 
computed from the points along each principal axis.}
\begin{center}
\begin{tabular}{|l|c|c|c|}
\hline
		&	Source axis	& Perp. axis	& Rotation axis	\\
\hline
Linear ($\mathcal{N}[\Psi] =0 $) &&&\\		
FD to KEH	&	0.25			&	0.24				&	0.10			\\
FD to ES	&	1.02			&	1.09				&	1.08			\\
ES to KEH	&	1.07			&	1.15				&	1.11			\\
\hline
$\mathcal{N}[\Psi] = \Psi^3$, $\lambda = 100 $ &&&\\
FD to KEH	&	3.56			&	3.44				&	0.21			\\
FD to ES	&	1.22			&	1.67				&	1.14			\\
ES to KEH	&	3.28			&	2.86				&	1.19			\\
\hline
$\mathcal{N}[\Psi] = \Psi^3$, $\lambda = -2.2$ &&&\\
FD to KEH	&	0.35			&	0.36				&	0.21			\\
FD to ES	&	1.96			&	2.02				&	2.08			\\
ES to KEH	&	1.77			&	1.99				&	1.92			\\
\hline
$\mathcal{N}[\Psi] = |\nabla\Psi|^2$, $\lambda = 3$ &&&\\
FD to KEH	&	1.73			&	1.47				&	1.33			\\
FD to ES	&	1.39			&	1.80				&	1.44			\\
ES to KEH	&	2.69			&	2.76				&	2.77			\\
\hline
$\mathcal{N}[\Psi] = |\nabla\Psi|^2$, $\lambda = -100$ &&&\\
ES to KEH	&	1.96			&	2.20				&	1.65			\\
\hline
$\mathcal{N}[\Psi] = \Psi\Box\Psi$, $\lambda =  -0.7$&&&\\
FD to KEH	&	0.53			&	0.48				&	0.20			\\
FD to ES	&	1.06			&	1.13				&	1.13			\\
ES to KEH	&	0.99			&	1.23				&	1.00			\\
\hline
\end{tabular}
\label{table:rmsdiff}
\end{center}
\end{table}

\section{Measuring range of convergence}
\label{sec:RConv}
For each code, a series of runs varying the nonlinear amplitude $\lambda$
determines the range of $\lambda$ for which the code gives a convergent
solution. The source distribution and boundary location are fixed.

Softening and continuation are used to extend the
maximum absolute value of  $\lambda$ that allows convergence.
For each nonlinear term, strikingly 
different behaviour is found for opposite signs of $\lambda$. 
That is, denoting by $\lambda_{\rm max}$ and $\lambda_{\rm min}$ the 
largest positive value and the smallest negative value of $\lambda$ for
which a code converges, we find for each nonlinear term that $\lambda_{\rm max}$ 
and $|\lambda_{\rm min}|$ differ by at least a factor of $100$.  
In each case, the sign of $\lambda$ that yields greatest convergence is 
opposite to the sign of the source term, where the source is large.   
The `favorable' sign damps the effect of the source distribution on the
field, while the `unfavorable' sign gives an amplified effective source. For
$\mathcal{N}[\Psi] = {\Psi^3,|\nabla\Psi|^2,\Psi\Box\Psi}$, the favorable signs of 
lambda are ${+,-,+}$, respectively.  

\begin{figure}[htp]
\begin{center}
\begin{tabular}{ccc}
(a)&(b)\\
\includegraphics[height=42mm]{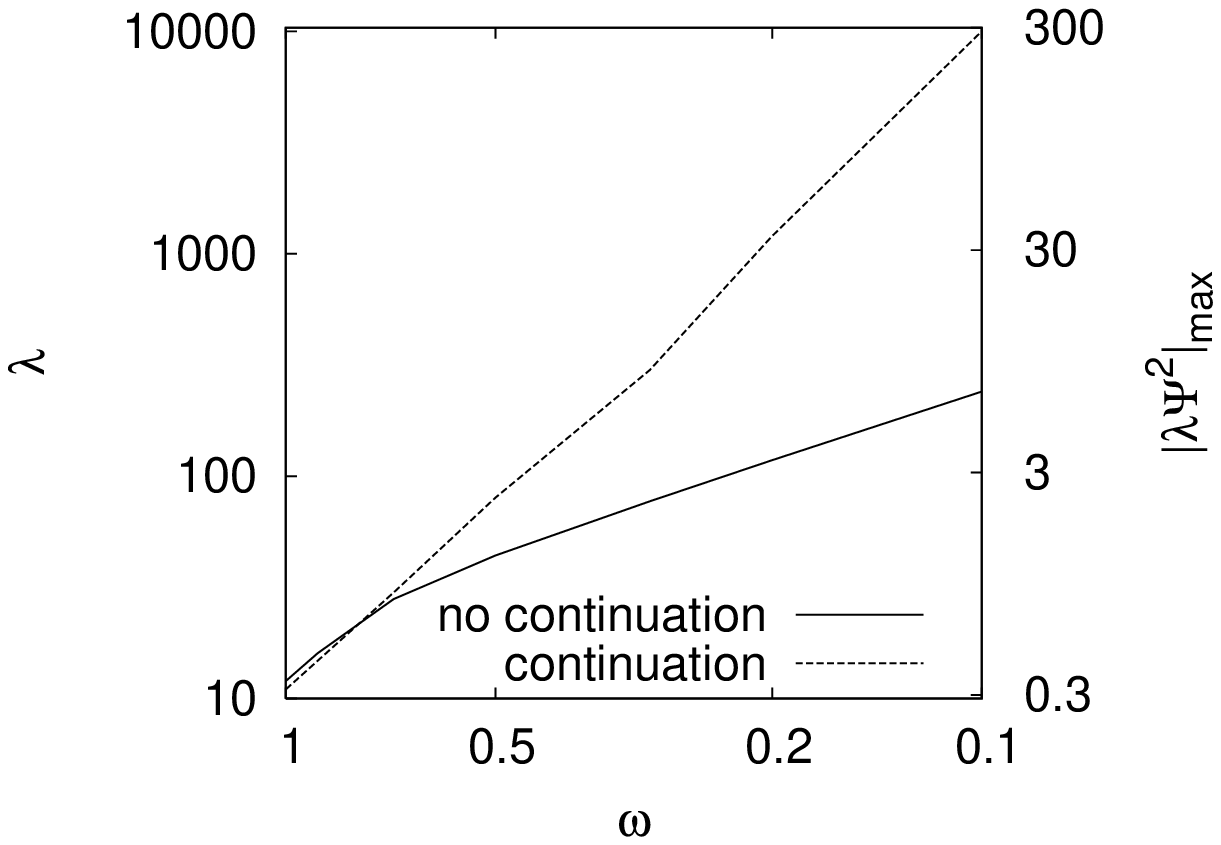} & 
\includegraphics[height=42mm]{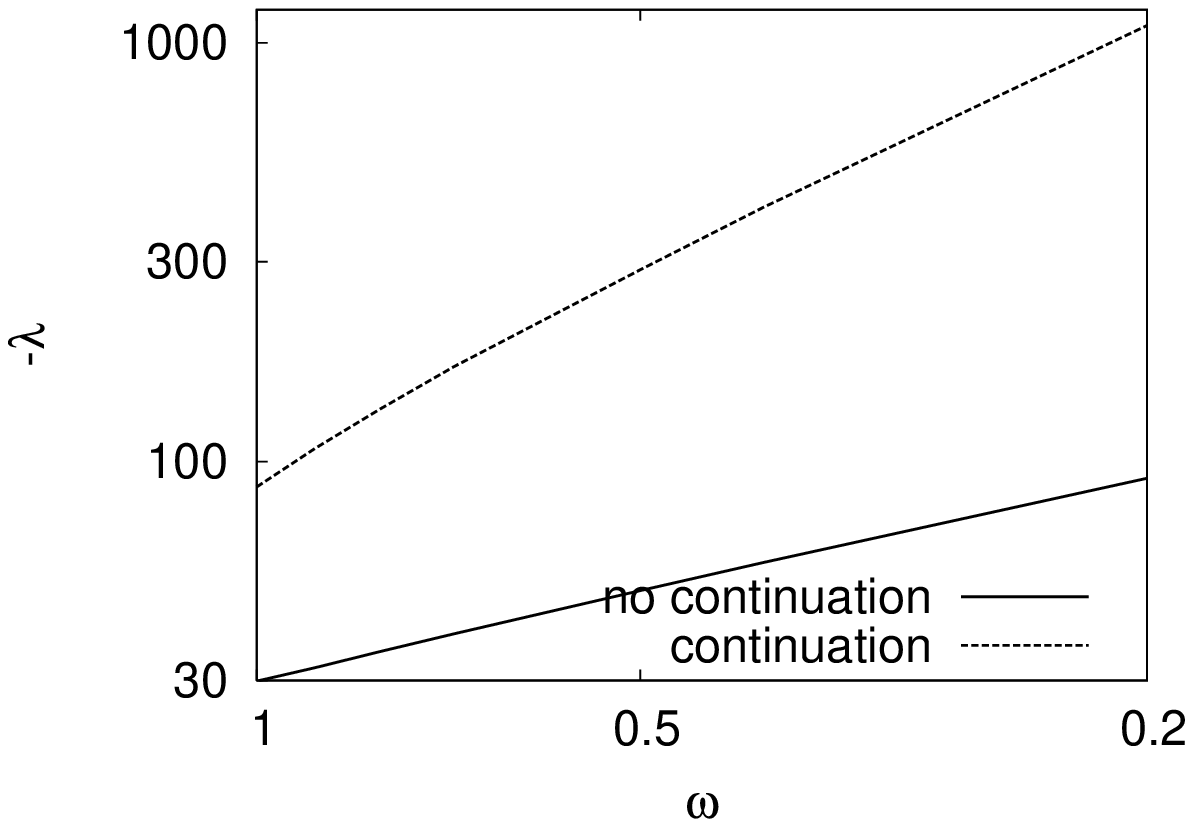} \\ 
(c)\\
\includegraphics[height=42mm]{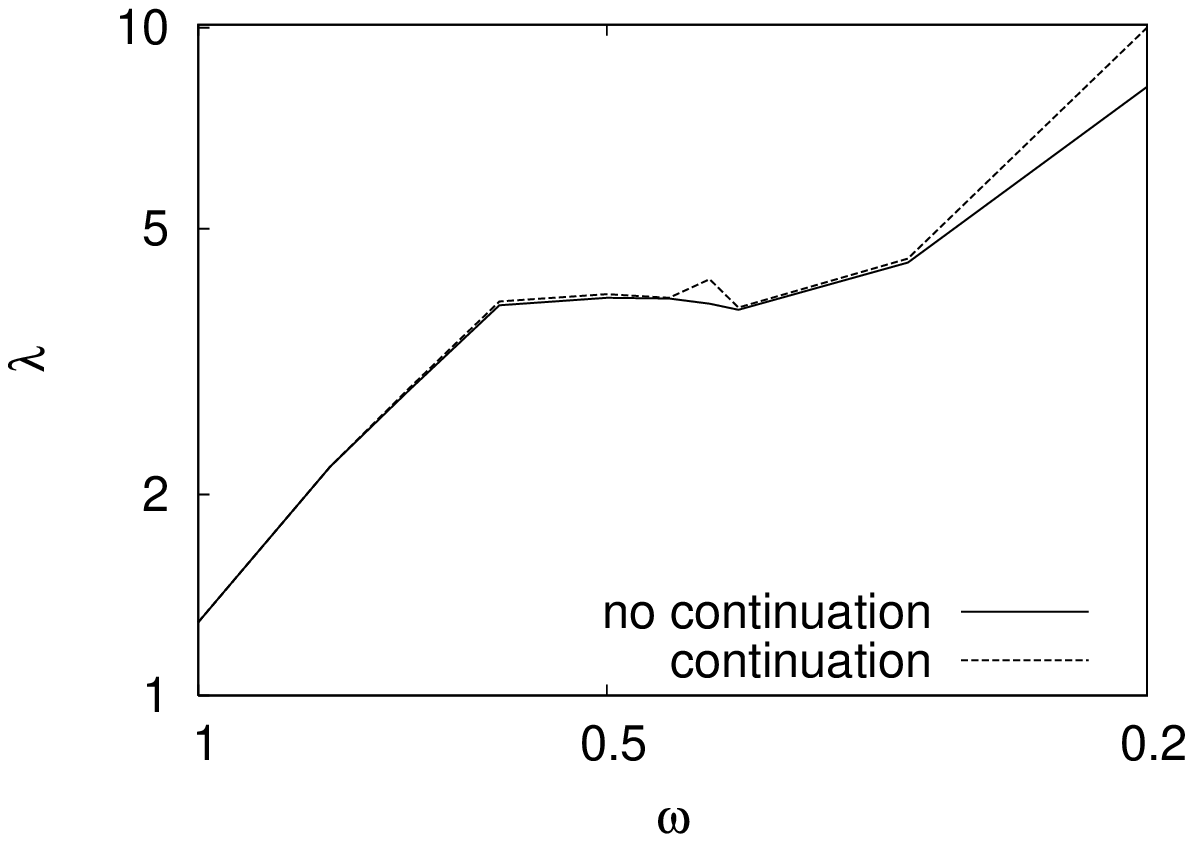} \\
\end{tabular}
\caption{The effect of a softening parameter $\omega$ and continuation
on range of $\lambda$. The plots indicate the limiting value of
$\lambda$ for convergence of the KEH code. Each panel corresponds to a
different nonlinear model, with $\mathcal{N}[\Psi]$ set to $\Psi^3$
(panel a), $|\nabla\Psi|^2$ (panel b), and $\Psi\Box\Psi$ (panel c).
In each case, the limiting $\lambda$ value of the opposite sign was
unaffected by softening or coninuation.}
\label{fig:softening}
\end{center}
\end{figure}

Softening and continuation improved the range of convergence for the
favorable sign, but had no effect on the limit of the
unfavorably signed $\lambda$. Figure \ref{fig:softening} shows the
effect of softening, parametrized by $\omega$, and of using continuation
on the limiting favorable-sign values of $\lambda$. 

Convergence is attained, when softening and continuation are
used, for all attempted favorable sign values for $\lambda$, except when
using the KEH method with the $\Psi\Box\Psi$ nonlinear term or when using the
Finite Difference method with the $|\nabla\Psi|^2$ nonlinear term.

The range of convergence results are given in Table \ref{table:converge}. If
softening and continuation maintain convergence to the largest tested value of 
$\lambda$, the uncertainty in the true maximum value or 
minimum value of 
$\lambda$ is indicated by $>$ or $<$.

\begin{table}
\caption{The range of nonlinear amplitude $\lambda$ in scalar models
for which codes converged. The first column gives the type of
nonlinearity, while the second column indicates the numerical method,
as discussed in the text.  The third and fourth columns, showing
$\lambda_{\rm min}$ and $\lambda_{\rm max}$, give the range of $\lambda$
for which convergence was achieved. Where an inequality is
given, no limiting value was found.}
\begin{center}
\begin{tabular}{|l|l|r|r|}
\hline
$\mathcal{N}[\Psi]$ & Code & $\lambda_{\rm min}$ & $\lambda_{\rm max}$ \\
\hline
$\Psi^3$ & KEH  &  -2.3 & 11  \\
$\Psi^3$ & KEH, softened/continued & -2.3 & $>$10000 \\
$\Psi^3$ & FD & -2.5 & $>$1000 \\
$\Psi^3$ & ES & -2.4 & $>$1000 \\
\hline
$|\nabla\Psi|^2$ & KEH &  -30 & 7.7\\
$|\nabla\Psi|^2$ & KEH, softened/continued & $<$-1000 & 7.7 \\
$|\nabla\Psi|^2$ & FD & -2.0 & 5.6 \\
$|\nabla\Psi|^2$ & ES & -360 & 7.3 \\
$|\nabla\Psi|^2$ & ES, continued & $<$-1000 & 7.3 \\
\hline
$\Psi\Box\Psi$ & KEH & -0.96 & 1.3 \\
$\Psi\Box\Psi$ &  KEH, softened/continued & -0.96 & 10 \\
$\Psi\Box\Psi$ &  FD & -1.8 & $>$1000 \\
$\Psi\Box\Psi$ & ES & -1.7 & $>$1000 \\
\hline
\end{tabular}
\label{table:converge}
\end{center}
\end{table}

\subsection{Boundary and convergence}
\label{sec:bdycvg}
As described in Sect.~\ref{sec:bns} below, 
Uryu has obtained a convergent helically symmetric 
BNS code within a region that extends about one wavelength from the center of the source, 
by using a waveless formulation outside that radius. Led by this
result, we explored the effects of the outer boundary placement on the range of
converging $\lambda$ for the fully helical code.

\begin{figure}
\begin{center}
\begin{tabular}{cc}
\includegraphics[height=42mm]{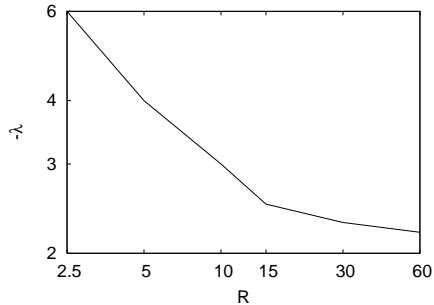}
\\
\end{tabular}
\caption{ The effect of the outer boundary $R$ on convergence in the
cubic nonlinear model ($\mathcal{N}[\Psi]=\Psi^3$).  The limiting value 
of $\lambda$ for which the KEH method converges as a function of outer 
boundary location, $R$, in units of orbital radii. }
\label{fig:outerbound}
\end{center}
\end{figure}

While bringing the boundary in has little effect on the limits of the favorably
signed $\lambda$, it does allow a significantly larger magnitude of the unfavorable
$\lambda$, as shown in Fig.~\ref{fig:outerbound}.

\section{Hybrid helical/waveless equations}
One expects a helically symmetric solution $\Psi$ to accurately approximate the outgoing 
solution only in the near zone, where the Coulomb part of the field is dominant.  That 
is, the two solutions agree when one can ignore the wavelike character of $\Psi$.  In 
that situation, one should be able to approximate the outgoing solution with comparable 
accuracy by using a waveless approximation in which only the Coulomb part of the field 
is retained.  We can obtain an approximation of this kind by replacing the D'Alembertian 
by the Laplacian -- by discarding second time derivatives.  In the last part (\ref{sec:compare}) 
of this Section, we compare results of this version of a waveless approximation to a solution 
that has exact helical symmetry.  We impose equivalent boundary conditions on the solutions 
by imposing helical symmetry only in the near zone, in effect matching to an exterior waveless solution.    
As we describe Sect. \ref{sec:bns}, in Uryu's neutron-star code, a helically symmetric near-zone 
is similarly matched on an initial hypersurface to an exterior waveless solution to the Einstein-Euler 
equations.  

We find as follows a helically symmetric near-zone $\Psi$ matched to an exterior waveless 
solution.  As before, after imposing helical symmetry, we have 
a scalar equation for the toy scalar model in the form
    \begin{equation}
      (\nabla^2-\Omega^2\partial_{\varphi}^2)\Psi = S + \lambda {\cal N}[\Psi],
    \end{equation}
where the effective source is a sum of the inhomogeneous source term and a 
nonlinear source. Instead of inverting
the operator on the left hand side in the equation with the wavelike 
Green's function ${\cal L}^{-1}$ of Eq. (\ref{eq:green}), we split the operator, writing
    \begin{equation}
      \nabla^2\Psi = S + \lambda {\cal N}[\Psi] +\Omega^2\partial_{\varphi}^2\Psi.
    \end{equation}
We regard the extra term on the right hand side as a
part of the effective source, and invert the Laplacian operator to solve
iteratively for $\Psi$. 

We can now truncate the extra source term
$\Omega^2\partial_{\varphi}^2\Psi$ (as in Uryu et al.\cite{WAT05}) at some
radius $R_{\rm trunc}$ written in units of the wavelength $\pi/\Omega$ of 
the quadrupole wave. The resulting solution is helically symmetric inside the
truncation radius and waveless outside. 

If this truncation radius is too large, the iteration fails to converge. In fact, 
this iterative method, in which we invert the Laplacian instead of the full second-order 
operator, fails to converge even for nonlinear terms that yield convergent solutions 
when we impose helical symmetry everywhere and use the iterative methods of the 
previous section.  For a given small truncation radius, the
critical value of $\lambda$ depends on the outer boundary radius $R_{\rm out}$, as in
the fully helical result of Figure \ref{fig:outerbound}. 

\begin{figure}[hbtp]
\begin{center}
\includegraphics[height=55mm]{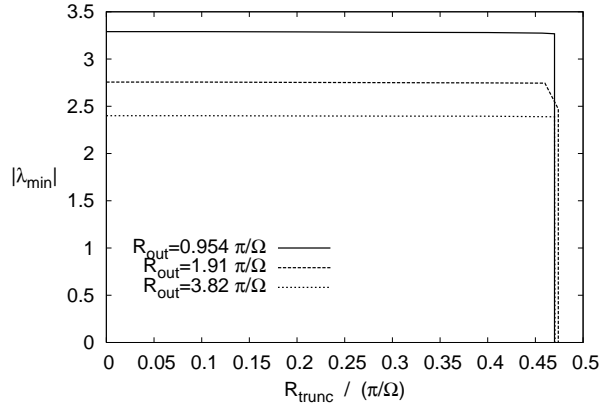}
\caption{Critical parameter $|\lambda_{\rm min}|$ is plotted as
 a function of $R_{\rm trunc}$. We plot curves for different
 computational outer boundaries $R_{\rm out}$. Note that $R_{\rm trunc}$
is normalized by $\pi/\Omega$.  The range of convergence is given for 
the iterative method described in this section with continuation used 
to reach a limiting value of $|\lambda_{\rm min}|$. 
}
\end{center}
\label{hybridcrlam}
\end{figure}
In Fig. \ref{hybridcrlam}, we plot the critical value $|\lambda_{\min}|$ of 
$\lambda$ as a function of 
$R_{\rm trunc}$, for the nonlinear term $\lambda\Psi^3$ with the unfavorable 
sign choice for $\lambda$. As in the models discussed above, the inhomogeneous source term is the double 
Gaussian, with $\Omega = 0.3$. In the upper panel, we
plot results of iteration without continuation. Different curves correspond to
different value of outer boundary radius, $R_{\rm out}$. As is clear from 
the figure, the value $\lambda_{\rm cr}=\lambda_{\rm min}$ is nearly independent of $R_{\rm trunc}$
up to maximum value of $R_{\rm trunc}$. Beyond that value, the iteration 
based on inverting the Laplacian does not converge for any finite value 
of $\lambda$. This transition
from convergence to non-convergence is abrupt, and we see a clear
limit, $R_{\rm trunc}\sim 0.47$, for the truncation radius. We
also see that using continuation we reached an absolute limit
of $R_{\rm trunc}\sim 0.47$ for any value of $R_{\rm out}$.

We note that the truncation radius of
the extra source (that is, $\Omega^2\partial_{\varphi}^2\Psi$) 
cannot be larger than a half of wave length of quadrupole component
of wave. This is roughly consistent with the results of Uryu's code described below.

\subsection{Comparing hybrid and waveless results}
\label{sec:compare}
\begin{figure}
\begin{tabular}{cc}
\begin{minipage}{.5\hsize}
\begin{center}
\includegraphics[height=42mm,clip]{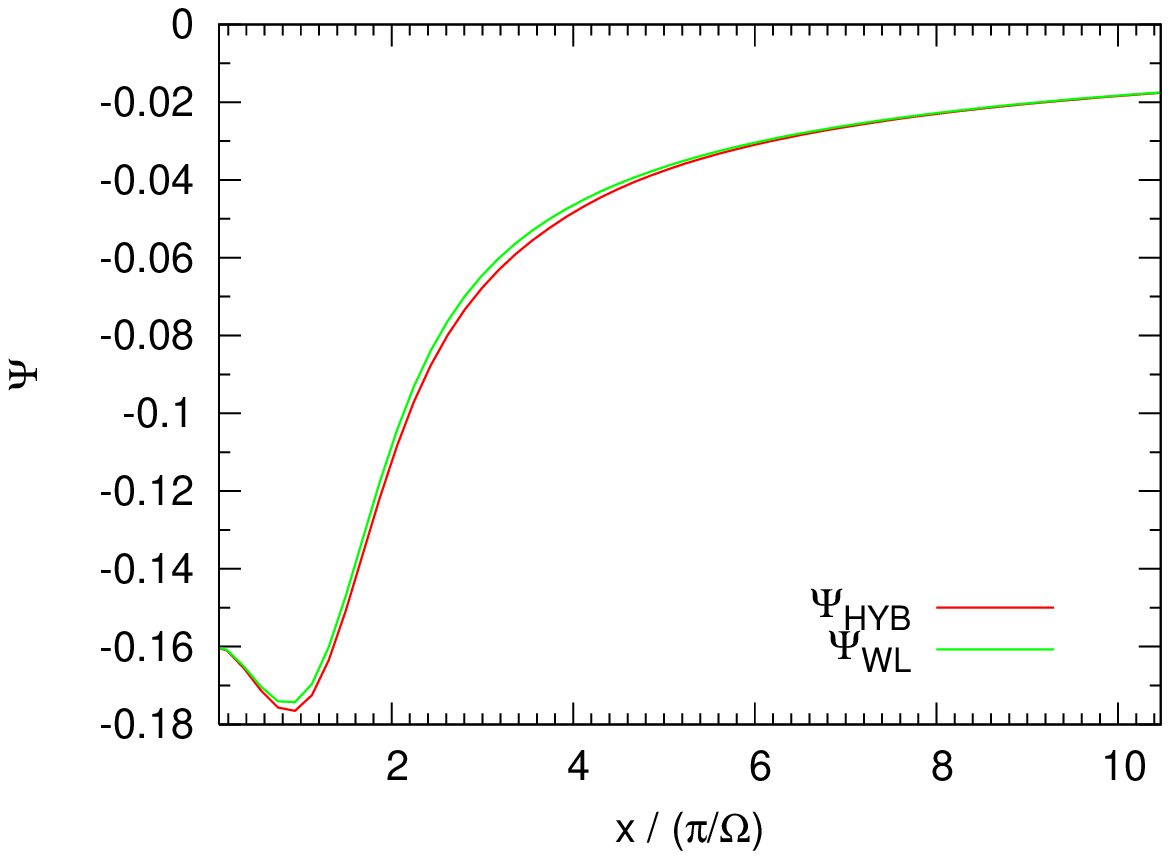}
\end{center}
\end{minipage} 
&
\begin{minipage}{.5\hsize}
\begin{center}
\includegraphics[height=42mm,clip]{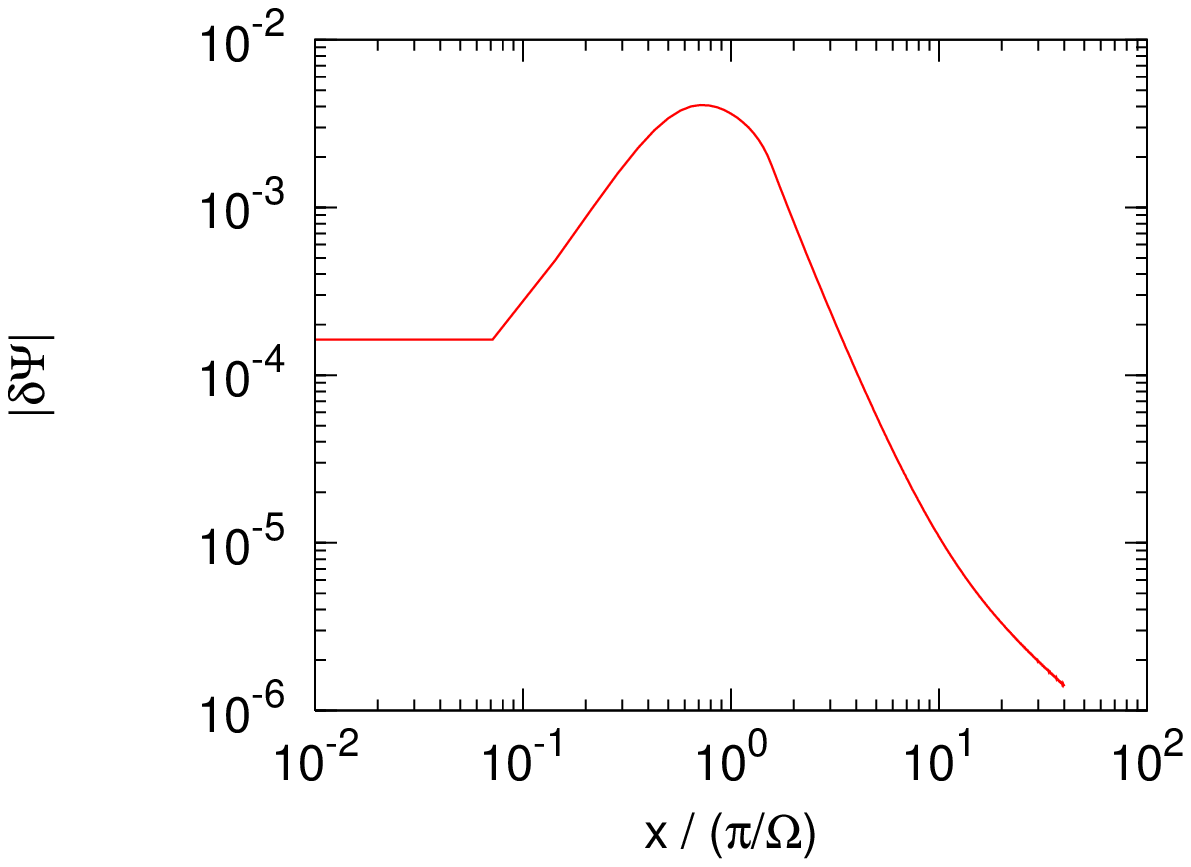}
\end{center}
\end{minipage} 
\end{tabular} 
\caption{
Left panel: 
Plot of hybrid ($\Psi_{HYB}$) and waveless ($\Psi_{WL}$) solutions 
along the source axis normalized by $\pi/\Omega$, the wavelength of the $l=m=2$
mode.  
Right panel:
The absolute error between hybrid and waveless solutions along the same axis.
}
\end{figure}
\label{fig:hybcompare}

In Fig. \ref{fig:hybcompare} we plot the difference between a hybrid solution
and a solution that is fully waveless.
The profile is along the source (along the x-axis).  
The resulting similarity of the solutions confirms our  
expectation that waveless and helically symmetric solutions should agree 
in the near zone. The level of disagreement, between 0.05 and 2\%, for $r$ less 
than one wavelength, appears primarily to result from the fact that the two 
solutions do not have identical boundary values at the boundary $R_{\rm trunc}$
of the region where one solution is helically symmetric and the other is waveless.
That is, by truncating the $\partial_\phi^2$ term in the effective source at 
$r=R_{\rm trunc}$, we match the helically symmetric solution to an exterior 
waveless solution, but to a slightly different waveless solution than that obtained 
by using the waveless equation everywhere.  


\section{Helically symmetric binary neutron star code}
\label{sec:bns}

In this section, we report the construction of a first helically 
symmetric code for binary neutron stars. The code modifies a recently developed 
numerical method for computing models of compact binary stars  in a waveless
approximation \cite{WAT05} to produce solutions to the full Einstein-Euler
system (the Einstein-perfect fluid equations)with exact helical symmetry.  
Waveless and helically symmetric solutions are expected to be accurate 
in the near zone, where the gravitational wave amplitude is small compared to
the Coulomb contribution to each metric potential; and in its present form, the
helically symmetric code converges only in the near zone.  Convergence of the
code is achieved by using the waveless formulation outside one wavelength.  In
effect, the outer waveless solution imposes boundary conditions on the
helically symmetric solution at the edge of the near zone.

\subsection{Formulation}

Our formulation is applied to Einstein's equation written in $3+1$ form, 
with spacetime ${\cal M} = \mathbb{R}\times\Sigma$.  
Let $n^\alpha$ be the future-pointing unit vector normal to the slices $\{t\}\times\Sigma$ of 
this foliation. 
The generator of the time coordinate $t^\alpha$ and the helical symmetry vector 
$k^\alpha = t^\alpha +\Omega\phi^\alpha$ are related 
to $n^\alpha$ and the shift $\beta^\alpha$ in the usual way,  
\beq
t^\alpha = \alpha n^\alpha + \beta^\alpha
\quad \mbox{and}\quad
k^\alpha = \alpha n^\alpha + \omega^\alpha,
\label{eq:tkvec}
\eeq
where $\omega^\alpha$ is a rotating shift vector 
$\omega^\alpha = \beta^\alpha + \Omega \phi^\alpha$.  

The projection 4-tensor 
$\gamma_{\alpha\beta}$ orthogonal to $n_\alpha$, defined by 
\beq
\label{eq:gamma}
 \gamma_{\alpha\beta} = g_{\alpha\beta} + n_\alpha n_\beta,
\eeq  
is associated with the 3-metrics $\gmabd(t)$ on the spatial 
slices $\Sigma_t$.  (Any spatial tensor on $\Sigma_t$ can be 
identified with a spacetime tensor orthogonal to 
$n^\alpha$ on all its indices).  
The metric then has the form,  
\beq
\label{eq:metric}
ds^2 = -\alpha^2 dt^2 +\gamma_{ij}(dx^i+\beta^i dt)(dx^j+\beta^j dt). 
\eeq

The extrinsic curvature of each slice $\Sigma$ is defined by 
\beq
\Kabd = -\frac12 \Lie_n\gmabd,  
\label{eq:Kab}
\eeq
where $\Lie_n$ operating on spatial tensors such as $\gmabd$ 
has the meaning 
\beq
\Lie_n \gmabd := \frac1{\alpha}\pa_t \gmabd - \frac1{\alpha}\Lie_\beta \gmabd,
\eeq
with $\pa_t \gmabd$ the pullback of $\Lie_t \gamma_\albe$ to $\Sigma$.
We introduce a conformal decomposition of the spatial metric, 
$\gmabd = \psi^4 \tgmabd$, and a condition 
$\tgamma = f$,  
where $f$ is the determinant of the flat metric $f_{ab}$ 
and $\tgamma$ the determinant of the conformal metric $\tgmabd$.  
Further $h_{ab}$ and $h^{ab}$ are introduced by 
\beq
\tgmabd=f_{ab}+h_{ab}\,,\quad  \tgmabu=f^{ab}+h^{ab}. 
\eeq

Helical symmetry,  \dis \Lie_k \gabd = 0$, 
implies for the 3-metric and extrinsic curvature, 
\beq
\Lie_k \gmabd = 0, \quad  
\Lie_k \Kabd = 0.  
\eeq
Using the relation ~(\ref{eq:tkvec}) between $n^\alpha$ and $k^\alpha$,
we have 
\beq
\Lie_n \gmabd = -\frac1\alpha \Lie_\omega \gmabd, \ \ \mbox{and}\ \ 
\Lie_n \Kabd = -\frac1\alpha \Lie_\omega \Kabd. 
\label{eq:Lien}
\eeq

Projections of the Einstein equation along the normal $n^\alpha$ 
are the Hamiltonian and momentum constraints,
\beqn
(\Gabd-8\pi\Tabd)n^\alpha n^\beta = 0, 
\label{eq:Hamcon}
\\
(\Gabd-8\pi\Tabd)\gmaa n^\beta = 0.  
\label{eq:Momcon}
\eeqn
while the spatial projection has trace and tracefree part 
\beqn
&&(\Gabd-8\pi\Tabd)\gamma^\albe = 0, 
\label{eq:tr}
\\
&&(\Gabd-8\pi\Tabd)\left(\gmaa\gmbb-\frac13\gmabd\gamma^\albe\right) = 0.
\label{eq:trfree}
\eeqn

The above set of equations are solved for the conformal factor $\psi$, 
the shift $\beta^a$, the lapse $\alpha$ and 
the deviation of the conformal metric from the flat metric $h_{ab}$, 
respectively, imposing coordinate conditions, the maximal slicing 
condition, $K=0$, and the generalized Dirac condition 
$\zD_b \tgmabu=0$, as in the waveless formulation \cite{SUF04}.  
In the following section, we concentrate on the treatment 
of the spatially tracefree part of Einstein equation (\ref{eq:trfree}),
solved for $h_{ab}$, in which the second time derivatives of $h_{ab}$ 
change the equation from elliptic (in the waveless formulation) to 
the mixed form that characterizes helically symmetric wave equations. 
A concrete treatment of the other components of the field equations 
that yield elliptic equations for $\psi$, $\beta^a$, 
and $\alpha$, as well as equations of matter source is given in Ref. \cite{SUF04}.  

The waveless code differs from the helically symmetric code by the requirement that  
$\tilde\gamma_{ab}$ have vanishing derivative along $t^\alpha$  
instead of vanishing derivative along $k^\alpha$. The extrinsic curvature and matter 
variables are still required to be helically symmetric.  These requirements 
result in elliptic equations for the field variables, including the non-conformal part of the 3-metric 
$h_{ab}=\tgmabd-f_{ab}$.  

In the helically symmetric code, with $\partial_t\tgmabd$ nonzero, the elliptic equations 
are replaced by Helmholtz equations for $h_{ab}$, whose source is almost 
identical to that of the waveless formulation.  
We implement a KEH solver for the helical formulation and investigate 
convergence of the iteration for a compact binary star source.  

%
\subsection{Helmholtz equation for $h_{ab}$}

Eq.~(\ref{eq:trfree}) has the form  
\beq
(\Gabd-8\pi\Tabd)\left(\gmaa\gmbb-\frac13\gmabd\gamma^{\alpha\beta}\right) = 
{\cal E}_{ab}-\frac13\gmabd\gmcdu{\cal E}_{cd}=0,
\label{eq:trfreeE}
\eeq
where 
\beq
{\cal E}_{ab}:= -\Lie_n \Kabd + \tR_{ab} + K\Kabd
-2 K_{ac}K_b{}^c -\frac1{\alpha}D_aD_b\alpha -8\pi S_{ab}, 
\label{eq:Eab}
\eeq
with $\tR_{ab}$ the Ricci tensor on $\Sigma$ 
associated with $\gmabd$, and $\Sabd$ the projection of 
the energy stress tensor, $\Sabd:=\Tabd\gmaa\gmbb$. 

By isolating the terms, $\Box h_{ab}:= (-\pa_t^2 + \zD^c\zD_c) h_{ab}$, 
that occur in ${\cal E}_{ab}$ in Eq.~(\ref{eq:Eab}), one can rewrite 
Eq.~(\ref{eq:trfreeE}) in the form 
\beq
\Box h_{ab} = {\cal S}_{ab}, 
\label{eq:boxh}\eeq
where $\Box$ is the flat d'Alembertian operator and $\zD^a := f^{ab}\zD_b$.  
Then, as in the scalar models, helical symmetry of the conformal metric, 
$\Lie_k \tgmabd = \Lie_k h_{ab}=0$, results in the operator 
\beq
\Box h_{ab} = -\pa^2_t h_{ab} + \zD^c\zD_c h_{ab} 
= (\zLap - \Omega^2\Lie^2_{\phi}) h_{ab}, 
\eeq
where the flat Laplacian is defined by $\zLap =\zD^c\zD_c$.  

Even when one uses the Cartesian components $h_{ij}$ of $h_{ab}$, however, 
$\zLap - \Omega^2\Lie^2_{\phi}$ does not coincide with the Helmholtz operator, 
because $\partial_\phi h_{ij}$ is a Cartesian component of ${\bm\phi\cdot\bf\zD} h_{ab}$, 
not of $\Lie_{\bm \phi} h_{ab}$. To isolate the Helmholtz operator 
${\cal L} = \zLap - \Omega^2\pa^2_{\phi}=\zLap-\Omega^2({\bm\phi\cdot\bf\zD})^2$, 
we write $\pa_\phi := \bm{\phi\cdot \zD}$ and find   
\beq
\Lie_{\bm \phi}^2\tgmabd
= \pa_\phi^2\tgmabd
\,+\,\left(\pa_\phi\tgamma_{ac}+\Lie_{\bm \phi}\tgamma_{ac}\right)\zD_b\phi^c
\,+\,\left(\pa_\phi\tgamma_{cb}+\Lie_{\bm \phi}\tgamma_{cb}\right)\zD_a\phi^c,
\label{eq:Liepa}
\eeq
where a relation $\phi^c\zD_c(\zD_a \phi^b)=0$ is used.  
Moving all terms in Eq.~(\ref{eq:Liepa}) except $\pa_\phi^2\tgmabd$ from 
the LHS to the RHS of Eq.~(\ref{eq:boxh}), we obtain the Helmholtz form
\beqn
{\cal L} h_{ab} = \bar{\cal S}_{ab}   
&:=& 
2\left(\bar{{\cal E}}_{ab} - \frac13 \tgmabd \tgamma^{cd}\bar{{\cal E}}_{cd}\right) 
\nonumber\\
&&-\frac13 \tgmabd \zD^e h^{cd}\zD_e h_{cd}
+\frac13 \tgmabd \Omega^2\pa_\phi h^{cd}\pa_\phi h_{cd}, 
\label{eq:Helm}
\eeqn
where the barred expression $\bar{\cal E}_{ab}$ is defined by 
\beqn
\bar{{\cal E}}_{ab}&:=& 
\Rnl_{ab} + \ttR_{ab}^{\psi}-\frac1\alpha D_a D_b\alpha
-2 \psi^4\tA_{ac}\tA_b{}^c  -8\pi S_{ab}
\nonumber\\
&&+\frac12\left(\frac{\psi^4}{\alpha^2}-1\right)\,\Omega^2\pa^2_{\phi}h_{ab}
+\, \frac{\psi^4}{\alpha^2}\,\Omega
\left(\Lie_{\phi}\Lie_{\beta}\tgmabd
+\frac12\Lie_{[\beta,\phi]}\tgmabd \right) 
\nonumber\\
&&+\,\frac{\psi^4}{2\alpha^2}\,\Omega^2
\left[\left(\pa_\phi\tgamma_{ac}+\Lie_{\bm \phi}\tgamma_{ac}\right)\zD_b\phi^c
\,+\,\left(\pa_\phi\tgamma_{cb}+\Lie_{\bm \phi}\tgamma_{cb}\right)\zD_a\phi^c
\right]
\nonumber\\
&&+\, \frac{\psi^4}{2\alpha^2}\,\Lie_{\beta}\Lie_{\beta}\tgmabd
\,+\,\frac{\psi^4}{\alpha}\tAabd\,\Lie_{\omega}\ln\frac{\psi^8}{\alpha}.
\label{eq:barEab}
\eeqn
In this expression for $\bar{\cal E}_{ab}$, 
the coordinate conditions $K=0$ and $\zD_b \tgmabu = 0$, mentioned above, are
imposed; and the tracefree part $A_{ab}$ of the extrinsic curvature, 
$ A_{ab}:=\Kabd - \frac13\gmabd K$, is introduced in the rescaled form 
$\tA_{ab}:=\psi^{-4} A_{ab}$.  
Terms $\Rnl_{ab}$ and $\ttR^\psi_{ab}$ arise from the conformal 
decomposition of the Ricci tensor
\cite{SUF04}.  

The source (\ref{eq:barEab}) can be written concisely 
without separating the second derivative term $\pa_\phi^2 h_{ab}$ explicitly 
as above.  Applying the helical symmetry condition (\ref{eq:Lien}) 
to Eq.~(\ref{eq:Eab}) and subtracting $\pa_\phi^2 h_{ab}$ term from 
both side of Eq.~(\ref{eq:trfreeE}), the source term of the exactly 
helical equation Eq.~(\ref{eq:Helm}) can be rewritten 
\beqn
\bar{{\cal E}}_{ab}&:=& 
\frac1\alpha\Lie_\omega (\psi^4 \tA_{ab})
- \frac12\Omega^2\pa^2_\phi h_{ab}, 
\nonumber\\
&&+\Rnl_{ab} + \ttR_{ab}^{\psi}-\frac1\alpha D_a D_b\alpha
-2 \psi^4\tA_{ac}\tA_b{}^c  -8\pi S_{ab}
\label{eq:barEab2}
\eeqn
which is equivalent to the above source term (\ref{eq:barEab}).

\subsection{A different formulation for the use of elliptic solver, 
and a comparison with the waveless approximation}
\label{sec:wateqs}

Instead of isolating the Helmholtz operator on the LHS for the case of helical symmetry, 
one can formally isolate an elliptic operator and solve the equation iteratively
using an elliptic solver as in the waveless approximation.  With this grouping 
of terms, the helical symmetry condition (\ref{eq:Lien}), applied to Eq.~(\ref{eq:Eab}), 
leaves the term $\Lie_\omega \Kabd$ as part of the effective source, and 
the Laplacian of $h_{ab}$ is separated out from $\tR_{ab}$.  
With gauge conditions $K=0$ and $\zD_b\tgmabu=0$, as before, we have 
\beq
\zLap h_{ab} = 
2\left(\widehat{{\cal E}}_{ab} - \frac13 \tgmabd \tgamma^{cd}\widehat{{\cal E}}_{cd}\right)
-\frac13 \tgmabd \zD^e h^{cd}\zD_e h_{cd}, 
\label{eq:wateq}
\eeq
where $\widehat{{\cal E}}_{ab}$ is given by 
\beq
\widehat{{\cal E}}_{ab}:= \frac1\alpha\Lie_\omega (\psi^4 \tA_{ab})
+ \Rnl_{ab} + \ttR_{ab}^{\psi} - \frac1\alpha D_a D_b\alpha
-2 \psi^4\tA_{ac}\tA_b{}^c - 8\pi S_{ab}.  
\label{eq:watEab}
\eeq
In Eq.(\ref{eq:watEab}), the first term appears instead of 
the last three lines in Eq.~(\ref{eq:barEab}).

We find that the helically symmetric code does not converge 
in this method when we solve the above set of equations on $\Sigma$ with 
a boundary that extends several wavelengths (or more) beyond the source.    
We were, however, able to obtain a converged solution when exact helical symmetry 
is imposed only in the near zone, within about a wavelength from the source, 
and the waveless approximation is used for larger $r$, effectively setting 
boundary conditions at the boundary of the helically symmetric inner zone.    
In a waveless approximation \cite{SUF04}, the time derivative of 
the conformal metric, $\pa_t\tgmabd$ is assumed to be zero.  As a result 
the extrinsic curvature is associated with the nonrotating shift $\beta^a$, 
\beq
\Kabd = - \frac12\Lie_n\gmabd = \frac1{2\alpha}\Lie_\beta\gmabd, 
\label{eq:Kabwat}
\eeq
instead of the rotating shift $\omega^a$ as in (\ref{eq:Lien}). 

In the next section, we compare the near-zone helically symmetric solution 
to a solution that is everywhere waveless.  For the near-zone helically symmetric 
solution, the change from helical symmetry to the waveless formulation at 
about one wavelength from the source implies for $\Kabd$ the condition 
\beq
\Kabd = \left\{
\begin{array}{ll}
\displaystyle
\frac1{2\alpha}\Lie_\omega\gmabd 
= \frac1{2\alpha}\left(\Lie_\beta\gmabd+\Omega \Lie_{\bm \phi}\gmabd\right), 
& \displaystyle
\mbox{for} \ \ r < f \frac{\pi}{\Omega}, \nonumber \\ \\
\displaystyle
\frac1{2\alpha}\Lie_\beta\gmabd + \frac1{3\alpha}\gamma_{ab}D_c(\Omega\phi^c), 
& \displaystyle
\mbox{for} \ \ r \geq f \frac{\pi}{\Omega},  \nonumber 
\end{array}
\right.
\label{eq:helicutoff}
\eeq
where $\pi/\Omega$ is the approximate wavelength of the $l=m=2$  
gravitational wave mode. The constant $f$, the coordinate radius of the 
helically symmetric zone in units of $\pi/\Omega$, is restricted 
to $f\alt 1$ for convergence.

\subsection{Numerical code}

Our numerical code is based on the finite difference code developed 
in Refs. \cite{UE00,WAT05}.  The code extends a KEH iteration scheme  
to the binary neutron star computation.  
Cartesian components of the field equations are solved numerically on 
spherical coordinate grids, $r$, $\theta$, and $\phi$.  
An equally spaced grid is used from the center of orbital motion to 
$5R_0$ where there are $16$ or $24$ grid points per $R_0$; 
from $5R_0$ to the outer boundary of computational region 
a logarithmically spaced grid has $60$ or $90$, 
where $R_0$ is the coordinate radius of a compact star along a line 
passing through the center of orbit to the center of a star.  
Accordingly, for $\theta$ and $\phi$ there are $32$ or $48$ 
grid points each from $0$ to $\pi/2$ and multipoles 
are summed up to $l=32$ \cite{UE00}.

\subsection{Numerical solution}

In this section we present results of our code for a binary system modeled by 
a perfect fluid having polytropic equation of state, $p=K\rho^\Gamma$, with 
$\rho$ the baryon density. We display results for the choices $\Gamma=2$, appropriate 
to neuton star matter; for compactness of a star in isolation $\compa=0.14 $; 
and for half the binary separation $d/R_0 = 1.375$.  
Solutions with helical symmetry are not uniquely specified by this choice of parameters 
and equation of state.  Because they are stationary solutions with equal amounts of 
ingoing and outgoing waves, they are not asymptotically flat, and one must find an 
appropriate choice of boundary conditions.  As discussed in the Consortium papers, 
one seeks conditions that minimize the amplitude of gravitational waves.  In toy 
models discussed above, conditions are fixed by the choice of a 
half-advanced plus half-retarded Green function at each iteration. In solving the 
Einstein equation, however, convergence is achieved only in the near zone, and, as 
we have emphasized, we impose boundary conditions by matching to an exterior waveless 
solution outside a coordinate radius $f\pi/\Omega$.   

\begin{figure}
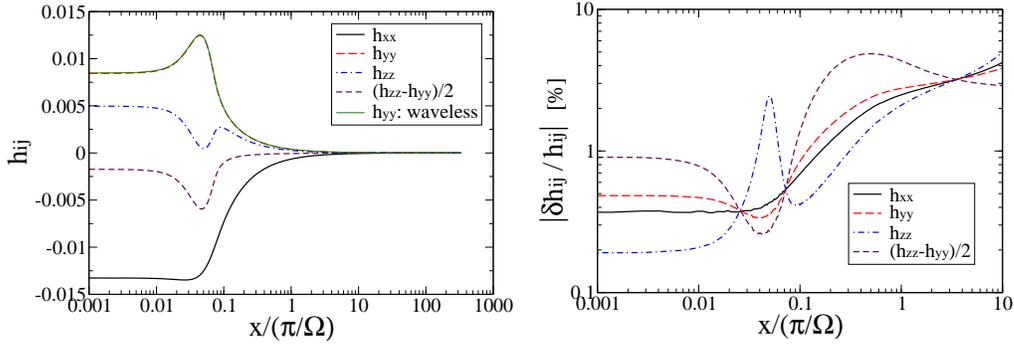

\begin{tabular}{cc}
\begin{minipage}{.5\hsize}
\begin{center}
\includegraphics[height=45mm,clip]{C014_plot_hij_Lap_he+wl.eps}
\end{center}
\end{minipage} 
&
\begin{minipage}{.5\hsize}
\begin{center}
\includegraphics[height=45mm,clip]{C014_plot_hij_error.eps}
\end{center}
\end{minipage} 
\end{tabular} 
\caption{
Left panel: 
Plot of components $h_{ij}$ along the $x$-axis 
normalized by $\pi/\Omega$, the wavelength of the $l=m=2$ mode.  
The solution to Eq.~(\ref{eq:wateq}) with the mixed helical and 
waveless source of Eqs.~(\ref{eq:watEab}) and (\ref{eq:helicutoff}), 
and the $h_{yy}$ component of the waveless solution are shown.  
Right panel: The fractional difference of these two solutions 
is plotted for selected components of $h_{ij}$ for $r < 10\pi/\Omega$.  
The fractional differences increase for larger $r$ because the values of 
the components $h_{ij}$ are themselves small.
A compact star extends from $x/(\pi/\Omega) = 0.0125$ to $0.079$, 
the boundary of computational region is set to 
$10^4 R_0 \sim 332\pi/\Omega$, and the cutoff constant $f$ in 
Eq.~(\ref{eq:helicutoff}) is set to $f=0.7$.
}
\label{fig:hab_Xaxis}
\end{figure}

For $f \alt 1$, the code converges, yielding a helically symmetric solution in the 
near zone $r\leq f\pi/\Omega$.  As shown in Fig.~\ref{fig:hab_Xaxis}, the solution 
is {\em nearly identical} to the waveless solution.  
The right panel shows a difference larger than 1\% only when the 
metric component itself is smaller than 0.03; as a percentage of $h_{ij}(r=0)$,
the difference is everywhere less than 1\%. The threshold of the value of $f$ for 
convergence is $0.7 \alt f \alt 1$ depending on the binary separation, compactness 
and resolution of finite differencing.  
We may expect from the result that, with boundary conditions 
that minimize the amplitude of gravitational waves, the exact helical solution 
will be close to the waveless solution near the source. This is a hopeful 
outcome: The waveless and helically symmetric formalisms 
are each intended to give a solution whose inaccuracy arises from neglecting gravitational waves, 
and they should give nearly identical results in the near zone where the 
gravitational wave amplitude is small compared to the Coulomb fields.    

%
%
%

In modeling binary neutron stars comparable accuracy is likely to result 
from codes that match a helically symmetric solution to a waveless solution, 
from a purely waveless code, and from a helically symmetric code. 
From a more mathematical perspective, however, finding a solution that has 
exact helical symmetry on the full spacetime is an appealing goal.  In the 
neutron star code, we have isolated a set of nonlinear terms that appear 
responsible for divergence outside the near zone.  
Improvement of the convergence of the code is required to extend the
matching radius beyond a few wavelengths, which may involve a use of
metric components in spherical instead of Cartesian coordinates.  Further 
investigation of this alternative is a subject of our future work.


\ack
This work was supported in part by NSF Grant PHY 0503366 and NASA Grant NNG05GB99G.
SY has been supported in part by Charles E. Schmidt College of Science
at FAU. We thank Chris Beetle, Richard Price, Eric Gourgoulhon, Francois Limousin, and 
Masaru Shibata for helpful discussions.

\end{document}